\documentclass{aa}  

\usepackage{natbib,twoopt,ulem}
\usepackage[breaklinks=true]{hyperref} 
\bibpunct{(}{)}{;}{a}{}{,}             
\makeatletter
  \newcommandtwoopt{\citeads}[3][][]{\href{http://adsabs.harvard.edu/abs/#3}%
    {\def\hyper@linkstart##1##2{}%
     \let\hyper@linkend\@empty\citealp[#1][#2]{#3}}}
  \newcommandtwoopt{\citepads}[3][][]{\href{http://adsabs.harvard.edu/abs/#3}%
    {\def\hyper@linkstart##1##2{}%
     \let\hyper@linkend\@empty\citep[#1][#2]{#3}}}
  \newcommandtwoopt{\citetads}[3][][]{\href{http://adsabs.harvard.edu/abs/#3}%
    {\def\hyper@linkstart##1##2{}%
     \let\hyper@linkend\@empty\citet[#1][#2]{#3}}}
  \newcommandtwoopt{\citeyearads}[3][][]%
    {\href{http://adsabs.harvard.edu/abs/#3}
    {\def\hyper@linkstart##1##2{}%
     \let\hyper@linkend\@empty\citeyear[#1][#2]{#3}}}
\makeatother

\usepackage[dvipsnames]{xcolor}

\definecolor{Mycolor2}{HTML}{00F9DE}

\newcommand{\BE}{\begin{equation}}
\newcommand{\EE}{\end{equation}}
\newcommand{\BA}{\begin{eqnarray}}
\newcommand{\EA}{\end{eqnarray}}
\newcommand{\Fig}[1]{Figure~\ref{fig:#1}}
\newcommand{\fig}[1]{Fig.~\ref{fig:#1}}

\newcommand{\eg}{{\it e.g.},}

\newcommand{\ie}{{\it i.e.}}

\usepackage{graphicx}
\usepackage{lscape}
\usepackage{rotating}
\usepackage{txfonts}
\usepackage{hyperref}

\begin{document} 

\title{A solar flare driven by thermal conduction observed in mid-infrared}

   \author{Fernando M. L\'opez\inst{1,4} \href{https://orcid.org/0000-0002-2047-6327}{\includegraphics[width=3mm]{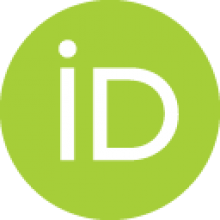}}
          \and
          C. Guillermo Gim\'enez de Castro\inst{1,2}
          \href{https://orcid.org/0000-0002-8979-3582}{\includegraphics[width=3mm]{orcid-ID.png}}
          \and
          Cristina H. Mandrini\inst{2}
          \href{https://orcid.org/0000-0001-9311-678X}{\includegraphics[width=3mm]{orcid-ID.png}}
          \and
           Paulo J. A. Sim\~oes\inst{1,5} 
           \href{https://orcid.org/0000-0002-4819-1884}{\includegraphics[width=3mm]{orcid-ID.png}}
          \and
          Germ\'an D. Cristiani\inst{2,3}
          \and
          Dale E. Gary\inst{6}
          \href{https://orcid.org/0000-0003-2520-8396}{\includegraphics[width=3mm]{orcid-ID.png}}
          \and
          Carlos Francile\inst{4,7}
          \href{https://orcid.org/0000-0002-9148-1780}{\includegraphics[width=3mm]{orcid-ID.png}}
          \and 
          Pascal D\'emoulin\inst{8,9}
          \href{https://orcid.org/0000-0001-8215-6532}{\includegraphics[width=3mm]{orcid-ID.png}}
          }

   \institute{Centro de R\'adio Astronomia e Astrof\'isica Mackenzie, Escola de Engenharia, Universidade Presbiteriana Mackenzie \\
Pr\'edio 45 T, cobertura, R. da Consolação, 896--7o andar--Consolação,
São Paulo, Brazil\\
               \email{guigue@craam.mackenzie.br}
         \and
         Instituto de Astronom{\'\i}a y F{\'\i}sica del Espacio, CONICET-UBA,  Ciudad Universitaria, Buenos Aires, Argentina
        \and
        Facultad de Ciencias Exactas y Naturales, Universidad de Buenos Aires, 1428 Buenos Aires, Argentina
        \and
         Facultad de Ciencias Exactas, F{\'\i}sicas y Naturales, Universidad Nacional de San Juan, Av. Jos\'e Ignacio de la Roza Oeste 590, J5402DCS, San Juan, Argentina 
         \and
        SUPA School of Physics and Astronomy, University of Glasgow, Glasgow G128QQ, UK
        \and
        Center for Solar-Terrestrial Research, New Jersey Institute of Technology, Newark, NJ 07102, USA
        \and
        Observatorio Astron\'omico F\'elix Aguilar (OAFA), Universidad Nacional de San Juan (UNSJ), San Juan, J5413FHL, Argentina
        \and
        LESIA, Observatoire de Paris, Université PSL, CNRS, Sorbonne Université, Université de Paris, 5 place Jules Janssen, 92195 Meudon, France
        \and
        Laboratoire Cogitamus, 75005 Paris, France
        }

   \date{Received ; accepted}

 
  \abstract
   {The mid-infrared (mid-IR) range has been mostly unexplored for the investigation of solar flares. It is only recently that new mid-IR flare observations have begun opening a new window into the response and evolution of the solar chromosphere. These new observations have been mostly performed by the  AR30T and BR30T telescopes that are operating in Argentina and Brazil, respectively. 
  }
   {We present the analysis of SOL2019-05-15T19:24, a GOES class C2.0 solar flare observed at 30~THz (10$\ \mu$m) by the ground-based telescope AR30T. Our aim is to characterize the evolution of the flaring atmosphere and the energy transport mechanism in the context of mid-IR emission.
   }
   {We performed a multi-wavelength analysis of the event by complementing the mid-IR data with diverse ground- and space-based data from the Solar Dynamics Observatory (SDO), the H--$\alpha$ Solar Telescope for Argentina (HASTA), and the Expanded Owens Valley Solar Array (EOVSA). Our study includes the analysis of the magnetic field evolution of the flaring region and of the development of the flare.
   }
   {The mid-IR images from AR30T show two bright and compact flare sources that are spatially associated with the flare kernels observed in ultraviolet (UV) by SDO. We confirm that the temporal association between mid-IR and UV fluxes previously reported for strong flares is also observed for this small flare. The EOVSA microwave data revealed flare spectra consistent with thermal free-free emission, which lead us to dismiss the existence of a significant number of non-thermal electrons. We thus consider thermal conduction as the primary mechanism responsible for energy transport. Our estimates for the thermal conduction energy and total radiated energy fall within the same order of magnitude, reinforcing our conclusions.
   }
   {}


\titlerunning{A solar flare driven by thermal conduction observed at mid-IR}
\authorrunning{F.M. L\'opez et al.}

   \maketitle
%
\section{Introduction}
\label{sect_intro}

The analysis of solar flares in mid-IR wavelengths, $ \lambda > 5\ \mathrm{\mu m}$, is a new area of research with only a few events reported in the literature thus far. \cite{Kaufmann13,Kaufmannetal:2015,Miteva16} and \cite{GimenezdeCastroetal:2018} used commercial IR cameras in the focus of small telescopes at room temperature with apertures ranging from 10 to 20~cm that produce images with diffraction limits around $15\arcsec$ and sensors centered at $\lambda=10\ \mathrm{\mu m}$ ($\nu = 30\ \mathrm{THz}$). These observations revealed that the mid-IR emission originates from compact regions and displays impulsive behavior. The only report of mid-IR flare spectral observations using Quantum Well Infrared Photodetectors (QWIP) cameras was from \citet{Pennetal:2016}, carried out at the 81~cm East Auxiliary branch of the McMath-Pierce telescope (now decommissioned), with  $\lambda=8.2\ \mathrm{\mu m}$ and $\lambda=5.2\ \mathrm{\mu m}$, \ie~37 and 58~THz, respectively, and diffraction limits on the order of $2\arcsec$. With this equipment, they were able to observe the mid-IR emission from two bright sources separated by about 13\arcsec during a C7-class flare in the classification of the Geoestationary Operational Environmental Satellite (GOES), with a spectrum compatible with optically thin thermal emission.

These studies have reported a good temporal or spatial agreement (or both) between mid-IR data and emission from other wavelengths, such as: microwaves (MW) \citep{Kaufmann13}, hard X-rays (HXR) \citep{Kaufmann13, Pennetal:2016}, white-light \citep{Kaufmann13,Pennetal:2016, GimenezdeCastroetal:2018}, and ultraviolet (UV) emission \citep{Miteva16}. Moreover, when compared to the flare emission in the submillimeter domain, the mid-IR presents higher flux density values. 

\citet{OhkiHudson:1975} first postulated that the mid-IR emission during flares should arise from an optically thin free-free source at chromospheric heights or from a blackbody spectrum from a heated portion of the photosphere. \citet{Heinzel12} used semi-empirical flare models, F1 \citep{Machadoetal:1980} and FLA \citep{Mauas90}, to calculate the flare spectrum from optical to radio wavelengths, and suggested that the mid-IR emission should be dominated by the free-free mechanism.

\citet{Trottet15}, in analyzing the data from \citet{Kaufmann13}, used the semi-empirical flare models developed by \citet{Machadoetal:1980} and the number of electrons and protons derived from hard X-Rays observed by {\sl Fermi} \citep{Meegan09} to compute the thermal emission produced at 30~THz, compared to the quiet Sun emission estimated from the semi-empirical model VAL \citep{VAL81}. 

\citet{Simoesetal:2017} employed radiative-hydrodynamic (RHD) simulations using RADYN \citep{Allred15} to compute the IR spectrum from synthetic flares arising from the dynamic atmosphere produced by precipitating electrons in the chromosphere. They show that the mid-IR spectrum originates in the mid-chromosphere from optically thin free-free radiation (up to $\sim50\ \mathrm{\mu m}$), and optically thick free-free at longer wavelengths, up to the cm-range. The enhanced free-free radiation is a result of the excess of free electrons due to the ionisation of hydrogen, and that the IR/sub-mm emission rapidly fades as the energy input ends, as hydrogen quickly recombines and depletes the excess of free electrons. In their simulations, \citet{Simoesetal:2017} note that no flare emission due to enhanced H$^-$ opacity arises from the photosphere.

The study of flares in the mid-IR can provide very important information regarding the evolution of H ionisation in the chromosphere to help understand how the energy is deposited in the solar atmosphere during flares. In this work, we analyze 30~THz images associated with a small GOES C2 class flare. Our interpretation of the observational data and analysis is discussed in Sect.~\ref{sec:Discussion} and, finally, our conclusions are presented in Sect.~\ref{sec:Conclusions}.  

\section{Observations}\label{sect_Obs}

The event SOL2019-05-15T19:24 (or SOL2019-05-15 for short) was classified as a C2.0 class flare in GOES X-ray flux. In soft X-rays (SXR) 
GOES band 1--8~\AA, its start time is 19:15~UT with maximum flux at 19:24~UT. The flare occurred in active region (AR) NOAA 12741, with heliographic coordinates N06W30 at 0:00 UT. Around 15 flares of B class\footnote{\url{https://www.spaceweatherlive.com/en/solar-activity/region/12741}} happened in the same AR since its appearance on the eastern solar limb, with the most intense being a B3.7 event on 12 May. The flare analyzed in this article is the strongest one during the AR disk transit from 7 May to 16 May. Starting with a description of the mid-IR instrument (see Sect. \ref{sec:mid-ir}), whose data motivated this study, we briefly enumerate the other instrumentation used as complementary support in the analysis (see Sect.~\ref{sec:comp-data}).     

\subsection{Mid-infrared}
\label{sec:mid-ir}

SOL2019-05-15 was observed at 30~THz ($10\ \mu$m) with the mid-infrared camera AR30T, installed at the Estaci\'on de Altura Carlos U. Cesco of the F\'elix Aguilar Astronomical Observatory (OAFA). The site is located in the Argentine Andes at an altitude of 2500~m above sea level. The camera is a FLIR model SC645, with an uncooled microbolometer array of 640$\times$480 pixels$^2$ and 17$\ \mathrm{\mu m}$ pixel size installed in the focus of a 20~cm Newtonian telescope. This telescope is "piggy backing"\ on the H-$\alpha$ Solar Telescope for Argentina \citep[HASTA,][]{Bagala99,FernandezBorda02}. The sensor spectral window ranges from 7.5$\ \mathrm{\mu m}$ ($\nu=40$~THz) to 13$\ \mathrm{\mu m}$ ($\nu = 23$~THz). While the telescope has a diffraction limit of 12.6\arcsec\ at the central frequency, its setup creates an oversampled pixel of 2.6\arcsec\ and parts of the solar limb are not accessible to the sensor. Data are acquired with a 1~s temporal cadence. However, to reduce the data storage requirements, in the absence of flares, only 1 every 10 frames is preserved. 

The mid-IR data are originally stored in a proprietary format that (during the reduction process) is transformed to the Flexible Image Transport System (FITS) format adding a World Coordinate System (WCS) header before its calibration in temperature. The process applied to the images includes: flat-fielding correction and alignment with the solar heliographic coordinates \citep{Manini19}. Additionally, a cross-correlation technique is applied to coalign the images and to minimize the jitter produced by atmospheric scintillation and telescope vibrations. 

\subsection{Complementary data} \label{sec:comp-data}
To understand the origin and evolution, as well as the energetics of the C2.0 flare, we complement our mid-infrared observations with data from the photosphere to the corona in various wavelengths. We use magnetograms and continuum images from the Helioseismic and Magnetic Imager \citep[HMI, ][]{Scherrer12} on board the Solar Dynamics Observatory \citep[SDO, ][]{Pesnell12}, H$\alpha$ data from HASTA, observations in microwaves from the Expanded Owens Valley Solar Array \citep[EOVSA,][]{Nitaetal:2016,Garyetal:2018}, and extreme UV (EUV) data from the Atmospheric Imaging Assembly \citep[AIA:][]{Lemen12} on board SDO. 

The analyzed SDO/HMI data consist of line-of-sight (LOS) magnetograms of AR 12741, selected from the full-disk data. We use magnetograms obtained with a temporal resolution of 45~s or 720~s, according to the analysis we perform. In all cases the magnetic data have a pixel resolution of around 0.5\arcsec. These magnetograms are used to study the evolution of the AR magnetic field, as described in Sect.~\ref{sec:magnetism} and to draw a scenario for the flare origin combining them with the EUV observations in Sect.~\ref{sec:Discussion}. We also use HMI pseudo-continuum images  that provide white-light (WL) information at a rate of one frame every 45~s. All images were corrected by the differential solar rotation effect. Base and running difference sequences were constructed to detect changes in brightness associated with the event. Additionally, light curves of the flaring region were made. 

The H$\alpha$ ($\lambda=656.3\ \mathrm{nm}$) data from HASTA have 1280$\times$1024 pixels with a spatial resolution of approximately 2\arcsec\ and are automatically taken in two different modes: normal (or patrol) and flare. In patrol mode, during routine observations, the telescope takes full disk images once every 90 seconds. When a flare begins, the instrument changes to flare mode and the temporal cadence increases to 5~s. The data set used here consists of images acquired in both modes, namely:\ patrol and flare (see Sect.~\ref{sec:chromos_EUV}).

After an upgrade in 2019 February, EOVSA now provides spectral images in the 1--18~GHz frequency range in spectral windows of 0.325~GHz bandwidth and with a time resolution of 1~s. The synthesized beams depend on the central frequency by the relationship $\sim 90\arcsec/\nu[\mathrm{GHz}]$. 

We mainly used SDO/AIA images at 1700~\AA, 1600~\AA, and 171~\AA\ (henceforth, AIA1700, AIA1600, AIA171). We selected, from the full disk images, subimages containing AR 12741 for the temporal range corresponding to the analyzed event. To follow the flare evolution, we co-aligned these images to compensate for solar rotation. The AIA images were obtained at a rate of one every 24~s for AIA1700 and AIA1600, and one every 12~s for AIA171. AIA images were processed to level 1.5 using the standard Solar Software (SSW) routines, conserving a pixel size of 0.6\arcsec. 

\section{Multi-wavelength data analysis}
\label{sec:phot-cor}

In this section, we describe the event and the environment in which it occurred at different levels in the solar atmosphere. We start with the magnetic evolution of AR 12741 and continue through chromospheric and coronal data analysis.

\subsection{Magnetic field evolution}
\label{sec:magnetism}

\begin{figure} 
\centering
\includegraphics[width = 0.4\textwidth]{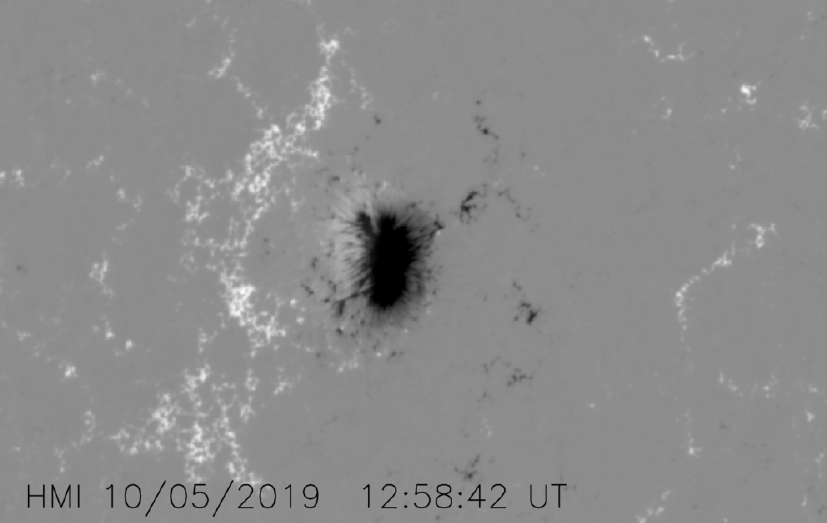}
\includegraphics[width = 0.4\textwidth]{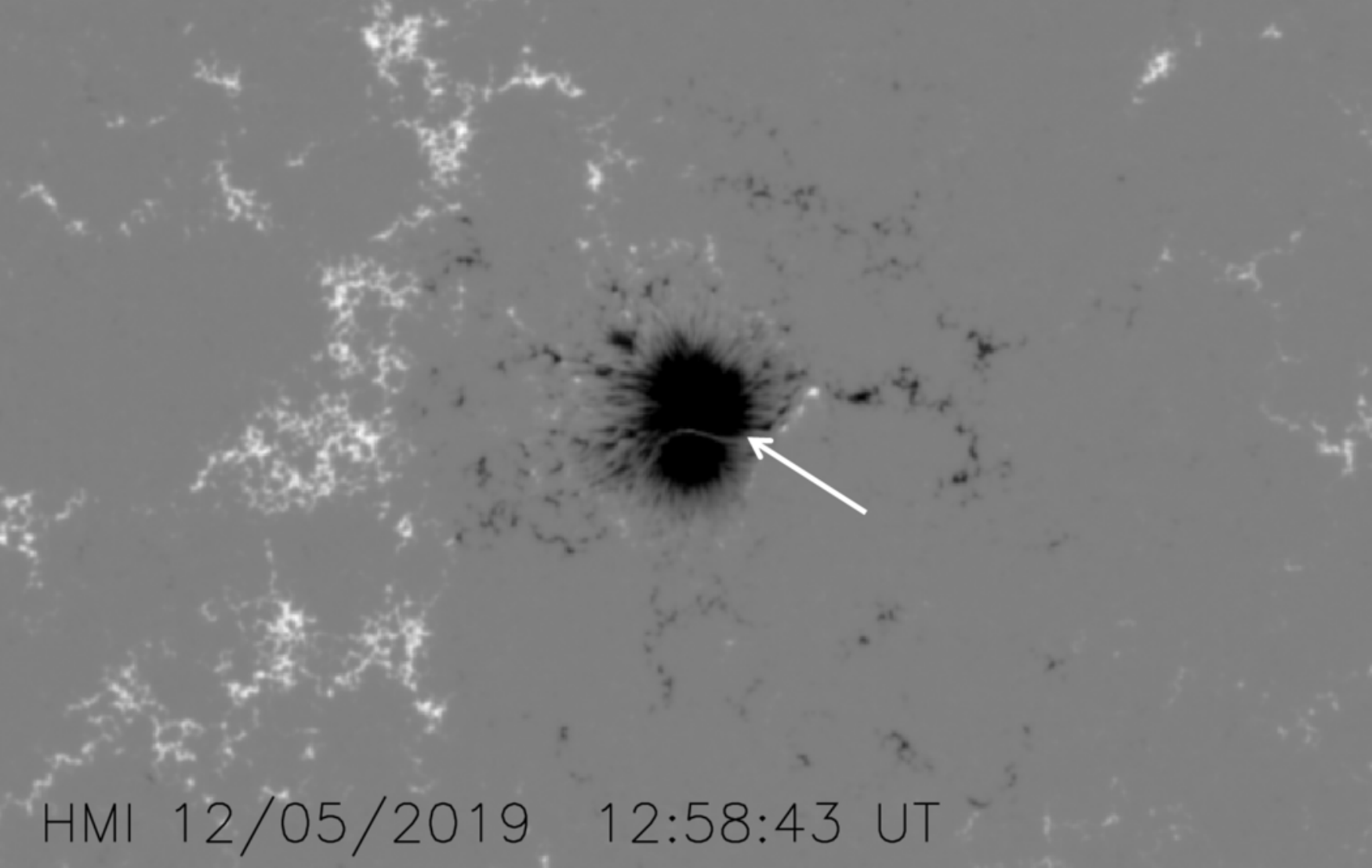}
\includegraphics[width = 0.4\textwidth]{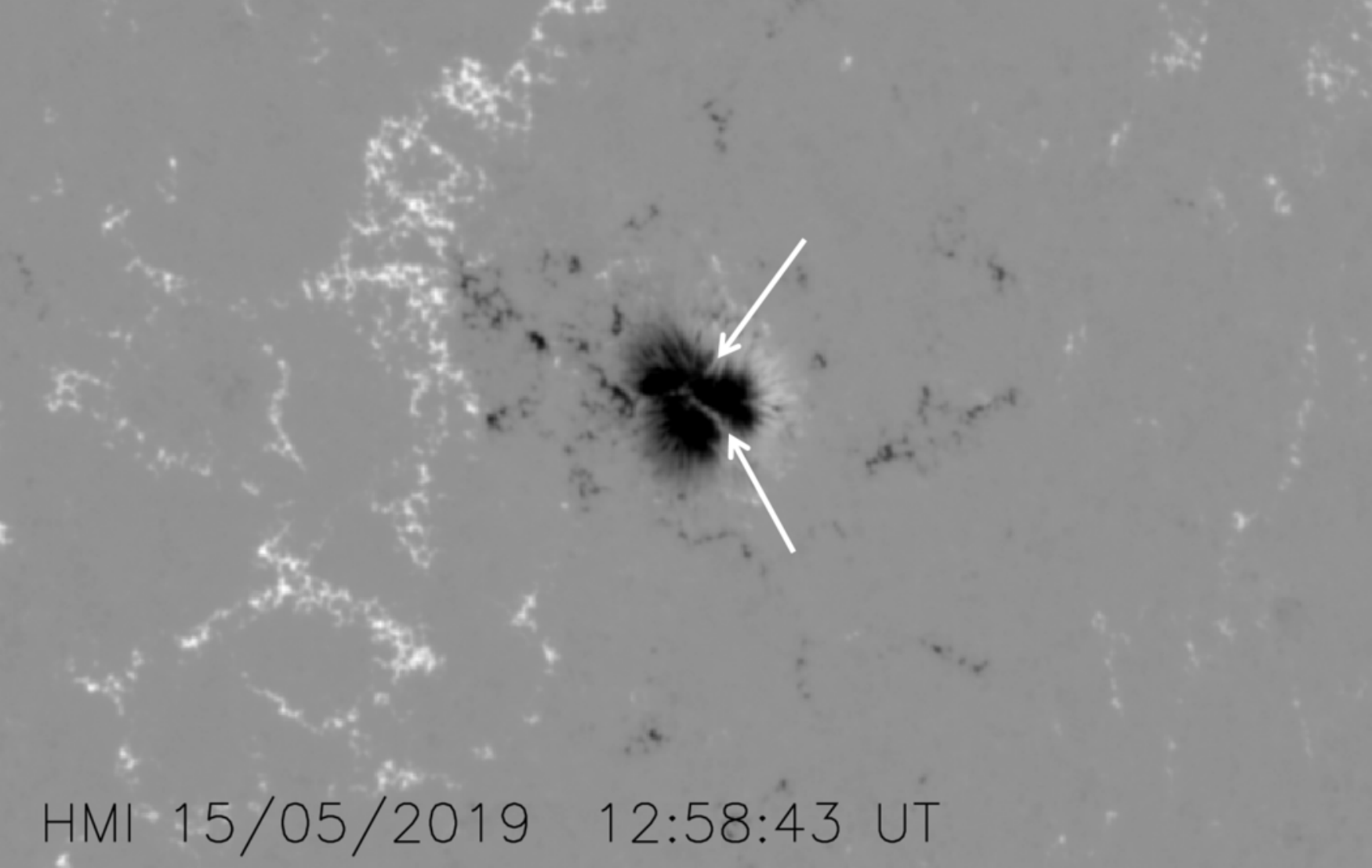}
\includegraphics[width = 0.4\textwidth]{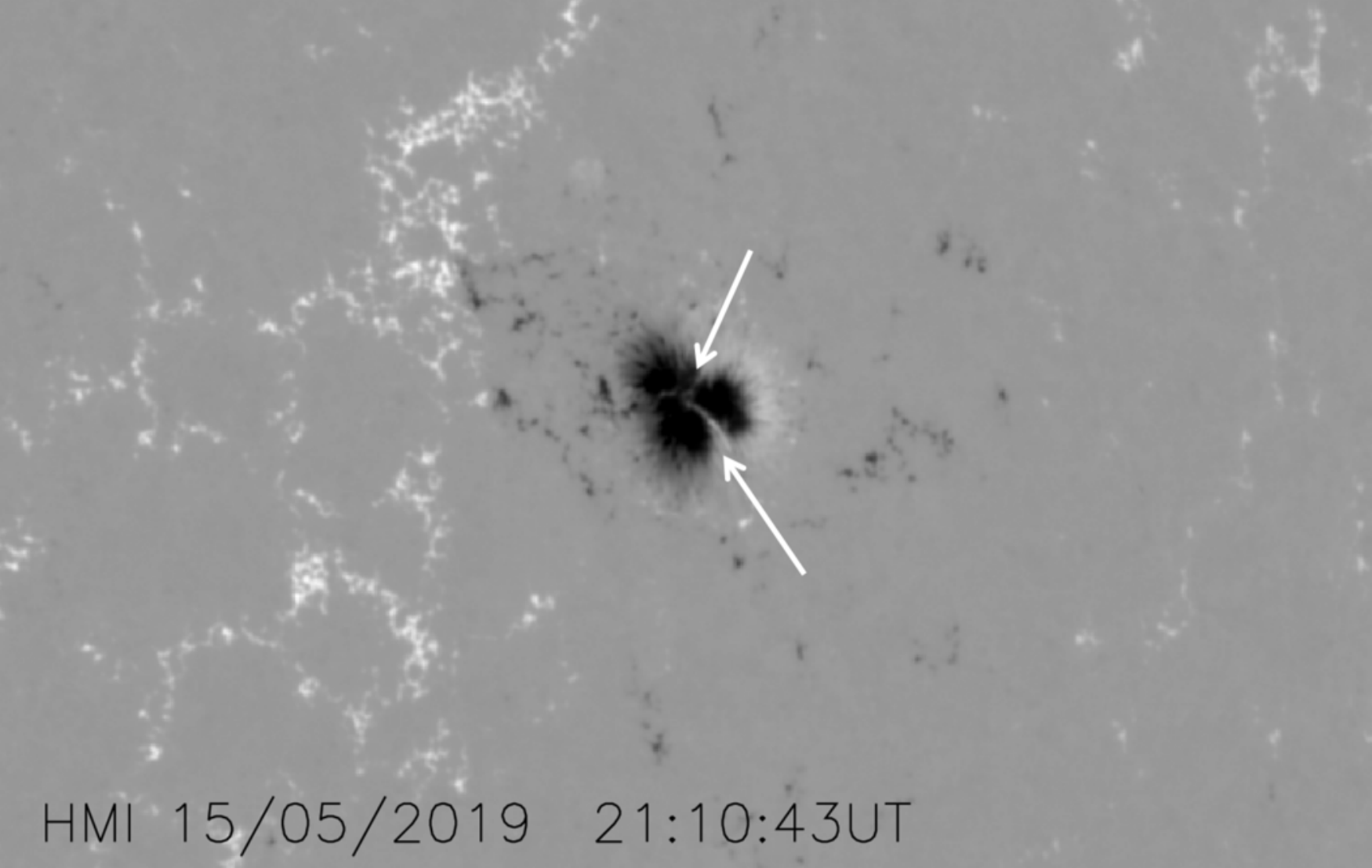}

\caption{SDO/HMI line-of-sight magnetograms of AR 12741, with 720~s time resolution, as it traverses the solar disk. 
The AR is formed by a compact and elongated negative polarity followed by a very disperse positive field. An east-west light bridge (see white arrow) is clearly observed on the second panel dividing the preceding spot into two sections. MMFs, formed by small bipolar fragments, are seen moving towards the moat region surrounding the spot. The AR is continuously decaying; by 15 May other light bridges are observed (as shown in the third and fourth panels marked by white arrows). The last two panels correspond to the flare day, in the early morning and at night. Black corresponds to negative magnetic field (pointing away from the observer) and white to positive field (towards the observer), the values of the field have been saturated above (below) 1000~G (-1000~G). The images are centered in the main negative polarity with horizontal and vertical sizes of 300\arcsec\ and 190\arcsec, respectively. The movie HMI-8-17May.mp4 complements this figure.
}
\label{fig:mag-fullevol}
\end{figure}

AR 12741 appeared on the eastern solar limb on 7 May 2019 as an already mature region formed by a preceding compact and elongated negative polarity and a very dispersed positive polarity. \Fig{mag-fullevol} illustrates different stages of the magnetic field evolution of AR 11741 from limb to limb.
This AR most probably emerged while being on the far side of the Sun. Ahead of its preceding polarity,  a very extended positive region of decaying magnetic field separates it from the other sizable AR (NOAA 12740) seen on the solar disk all along its solar disk transit\footnote{\url{https://www.solarmonitor.org}}. From its appearance and until its occultation on the western limb, the preceding sunspot was observed to be decaying. Small magnetic flux fragments were seen to move radially outwards from the spot all during this period. These bipolar features, called "moving magnetic features" \citep[MMFs:][]{Harvey73}, moved in the moat region away from the sunspot.  This moat region surrounds the spot with an annular shape, starting from the spot boundary and ending at the border of positive supergranular cells to the east and the negative ones to the west (\fig{mag-fullevol}). No other changes such as flux emergence episodes were observed during the AR transit, except this constant advection of small flux fragments by the moat flow that mark the decay phase of the sunspot. By 12 May at $\sim$01:00 UT a clear light bridge is seen in HMI magnetograms traversing the sunspot in the east-west direction, dividing it into two fragments. This light bridge is already seen in WL images by the end of 10 May\footnote{\url{https://www.helioviewer.org}}. By early 15 May the preceding sunspot appears in HMI magnetograms to be divided into three sections because of the appearance of other light bridges; this is already visible on the previous day in WL images. The evolution just described can be followed in the HMI magnetic field movie (see HMI-8-17May.mp4) attached to Fig.~\ref{fig:mag-fullevol}.

Though the AR is frankly decaying, one solar rotation later at the corresponding location on the solar disk one can still observe a similar magnetic flux pattern as that of AR 12741 and the preceding AR 12740. This is evident when comparing the full disk images on 12 May 2019, when AR 12741 was located at the disk center, to that of 8 June 2019 one solar rotation later. This is coherent with the well-known result that ARs and, in particular sunspots, can survive from days to several weeks \citep{vanDriel15}.

\begin{figure}
\begin{center}
\begin{tabular}{c c}
\includegraphics[width=0.23\textwidth]{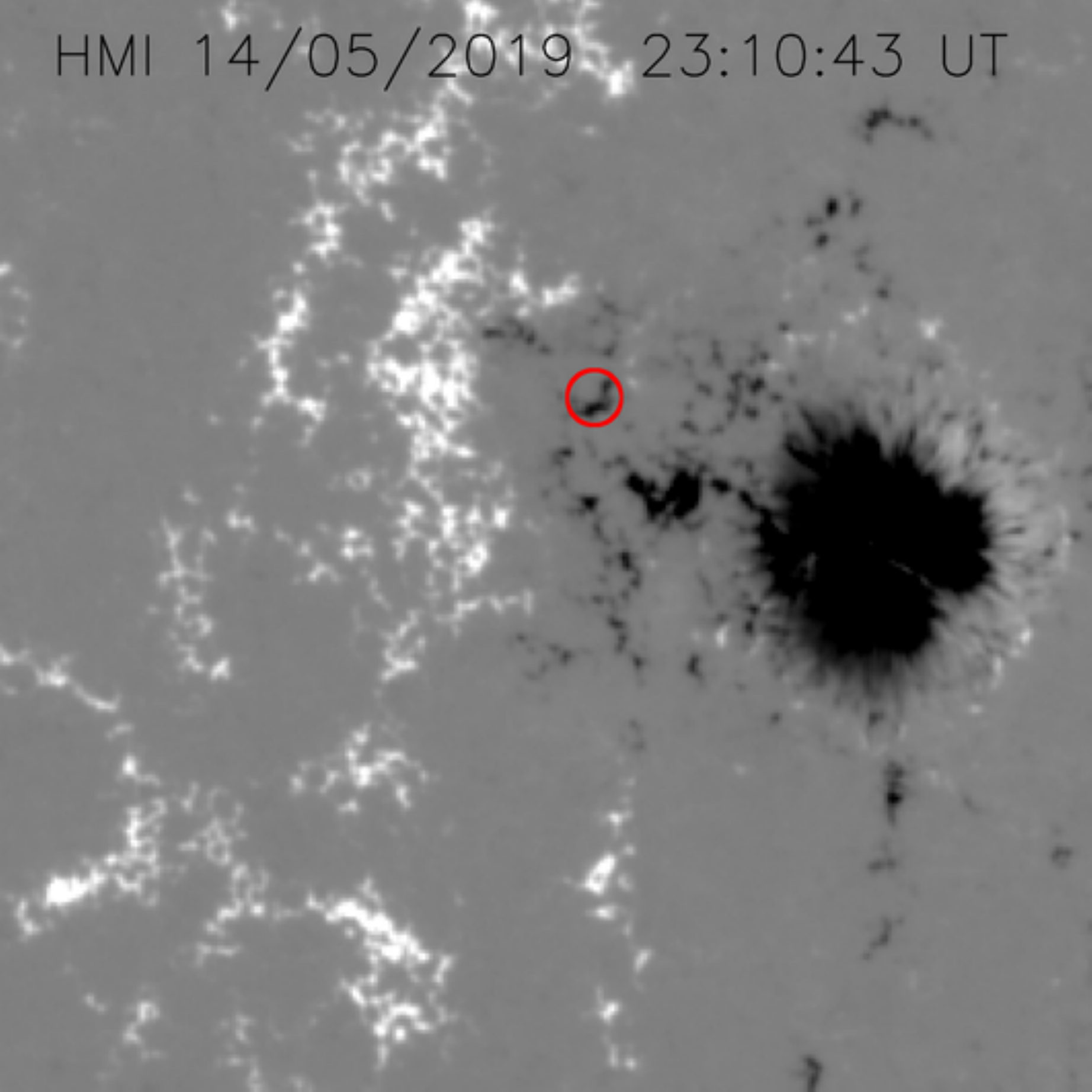}\hspace{-0.38cm}\vspace{-0.12cm} &
\includegraphics[width=0.23\textwidth]{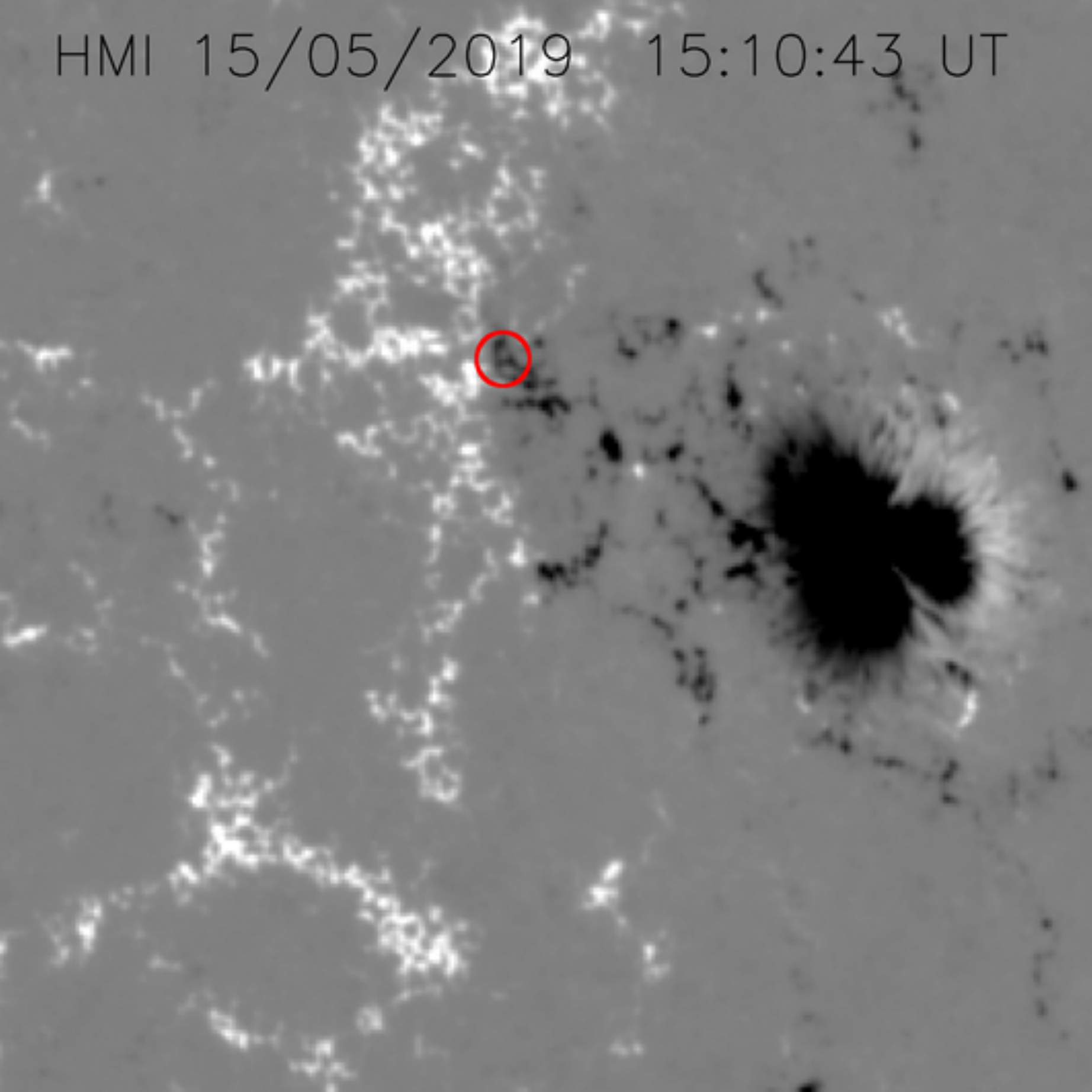} 
\end{tabular}
\begin{tabular}{c c}
\includegraphics[width=0.23\textwidth]{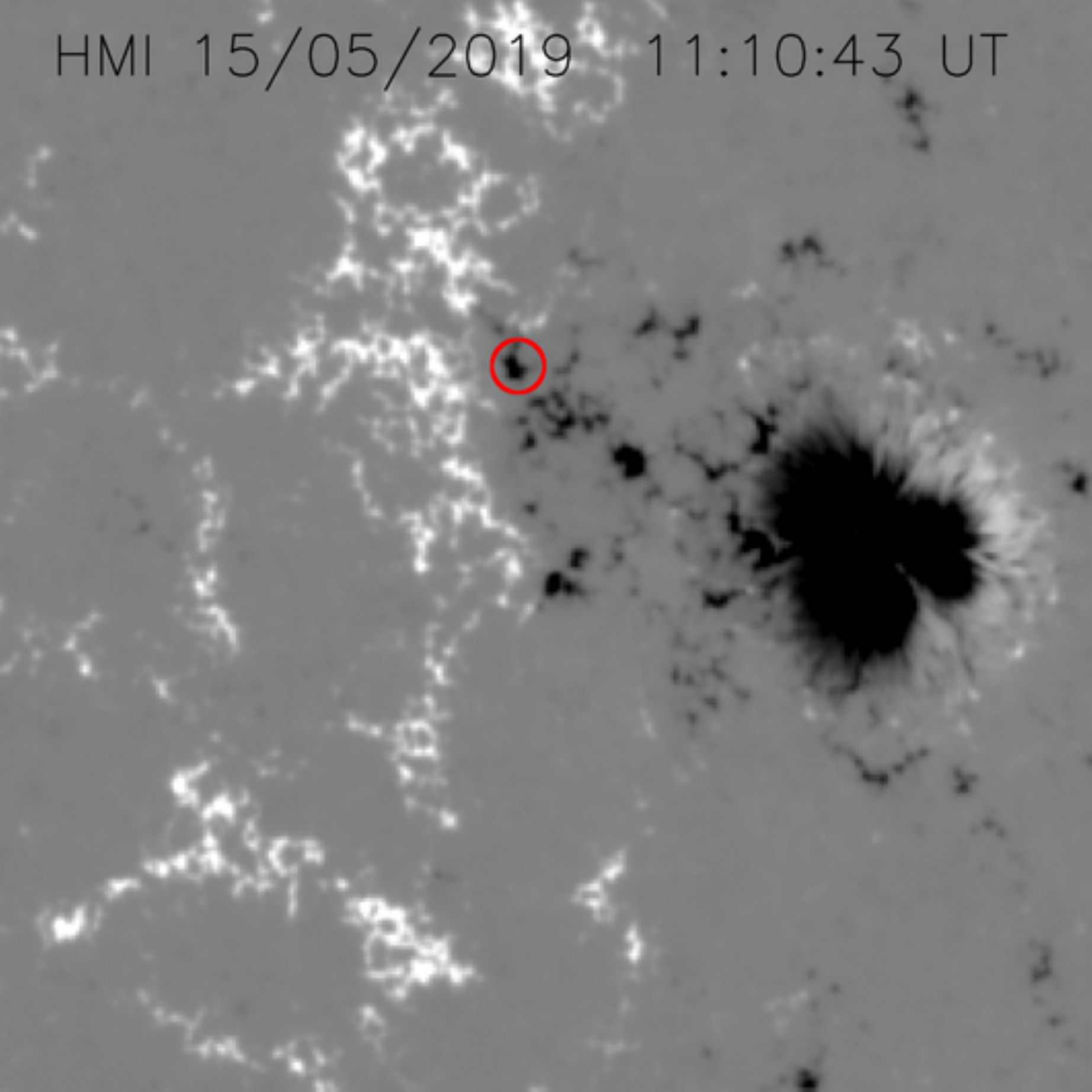}\hspace{-0.38cm}\vspace{-0.12cm} & 
\includegraphics[width=0.23\textwidth]{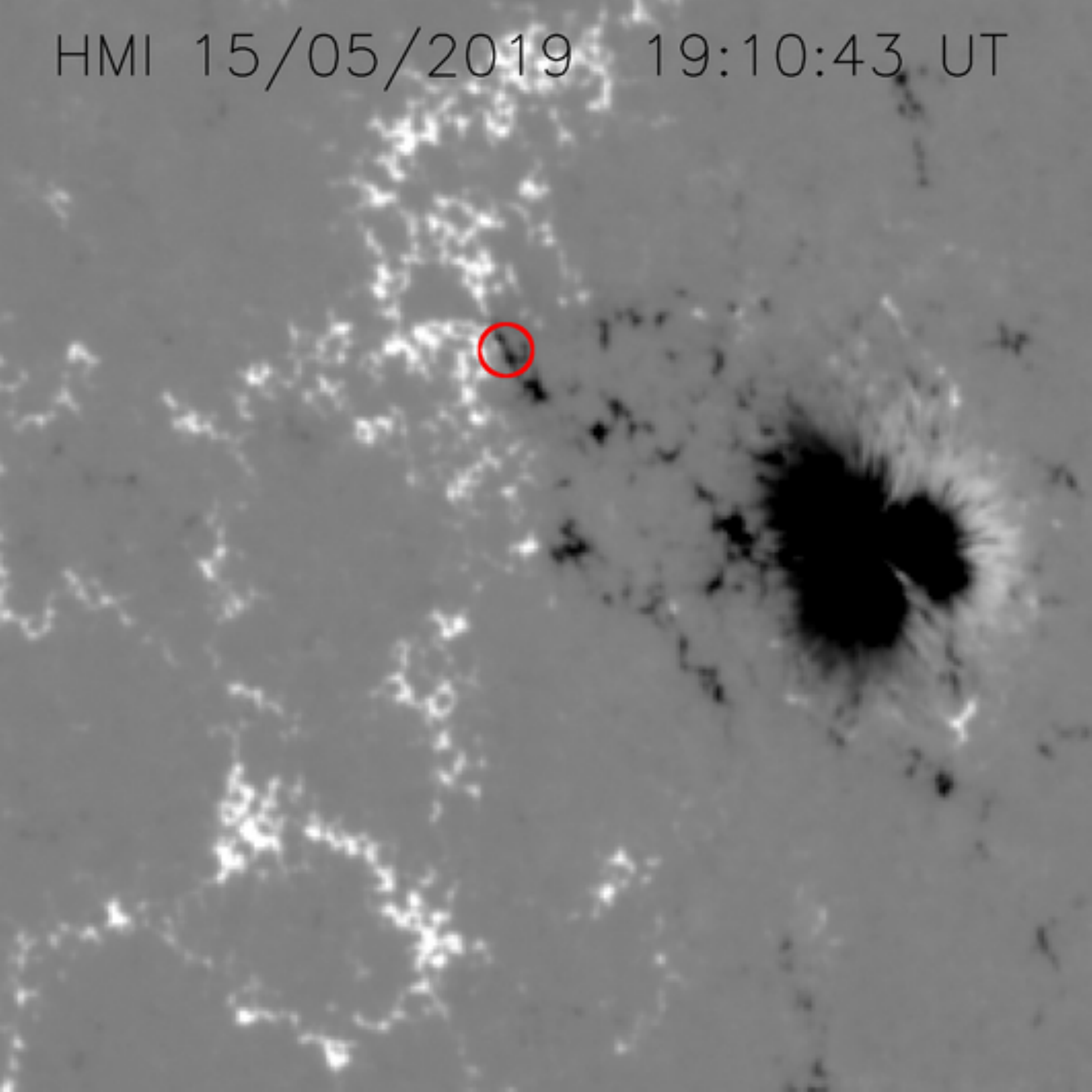}
\end{tabular}
\caption{SDO/HMI line-of-sight magnetograms of AR 12741 on 15 May 2019 with a 45~s time resolution. The red circles mark the small negative feature that we track to compute the moat flow speed at different times. Its evolution indicates that as it moves closer to the positive field its flux disperses and decreases. The convention for the field intensity is the same as in \fig{mag-fullevol} and the images have a horizontal and vertical sizes of 150\arcsec. The movie HMI-15May.mpeg complements this figure.
}
\label{fig:mag-arevol}
\end{center}
\end{figure}

The panels in \fig{mag-arevol} display the evolution of the line-of-sight magnetic field on the flare day using maps with 45~s time resolution. These maps, with the highest available temporal resolution, are used to estimate the speed of the moat flow. We identify a small negative polarity feature and we track it by eye moving out from the preceding polarity to the east. On 14 May at 17:10 UT, this feature was located at 7.5~Mm from the negative polarity border at -1000~G of the sunspot and it reached the positive disperse field (at 23.5~Mm from the same location) on 15 May at 19:10 UT (see the movie HMI-15May.mpeg). The moat flow speed computed after correcting for projection effects ranges from $\approx$460~m~s$^{-1}$ to $\approx$170~m~s$^{-1}$ during this time period. These values are well within the values measured in other cases using HMI data \citep[see e.g.,][]{Lohner-Bottcher13}. 

\begin{figure}
\centering
\includegraphics[width=8.5cm]{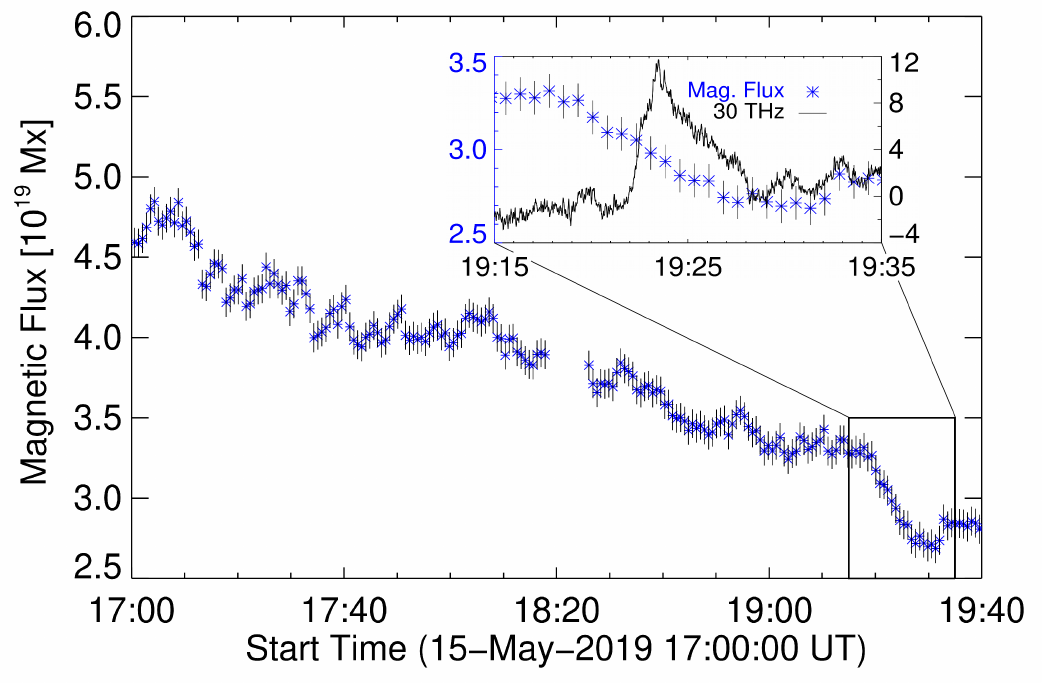}
\caption{Evolution of the absolute value of the negative magnetic flux in the region surrounded by the red square shown in the first panel, last row, of \fig{mapsAR}. Computations are done for values of the field below $-10$~G and the error bars are calculated considering a magnetic field error of 5~G. The time evolution between 19:15 and 19:35 UT, is zoomed in the upper-right corner and displayed together with the temporal evolution of the excess brightness temperature in 30~THz (see \fig{curves} and Sect.~\ref{sec:midIR-evol} for details about mid-IR brightness temperature calculation).} 
\label{fig:flux-evol}
\end{figure}

Continuous magnetic flux cancellation was observed as the negative fragments moved away from the sunspot and approached the positive following AR polarity. We have estimated the variation of the negative flux at the location of one of the main flare kernels on the negative field (see the red square in the first panel of the last row in \fig{mapsAR}), 
which we envision could be the footpoint of one of the loops involved in the flare as follows from the description in Sect.~\ref{sec:Discussion}. \Fig{flux-evol} shows the clear decreasing trend of the negative flux, which starts before the flare. This decrease becomes steeper at around 2 minutes before the impulsive phase of the flare, as shown by the inset at the top right of this figure depicting the mid-IR emission (see Sect.~\ref{sec:midIR-evol},
and flattens immediately after. Due to the broad distribution of the positive field, it is very difficult to identify and measure the equivalent positive flux decrease.

\subsection{H$\alpha$ and EUV evolution}
\label{sec:chromos_EUV}

\begin{figure*}[ht!]
\centering
\includegraphics[width=18cm]{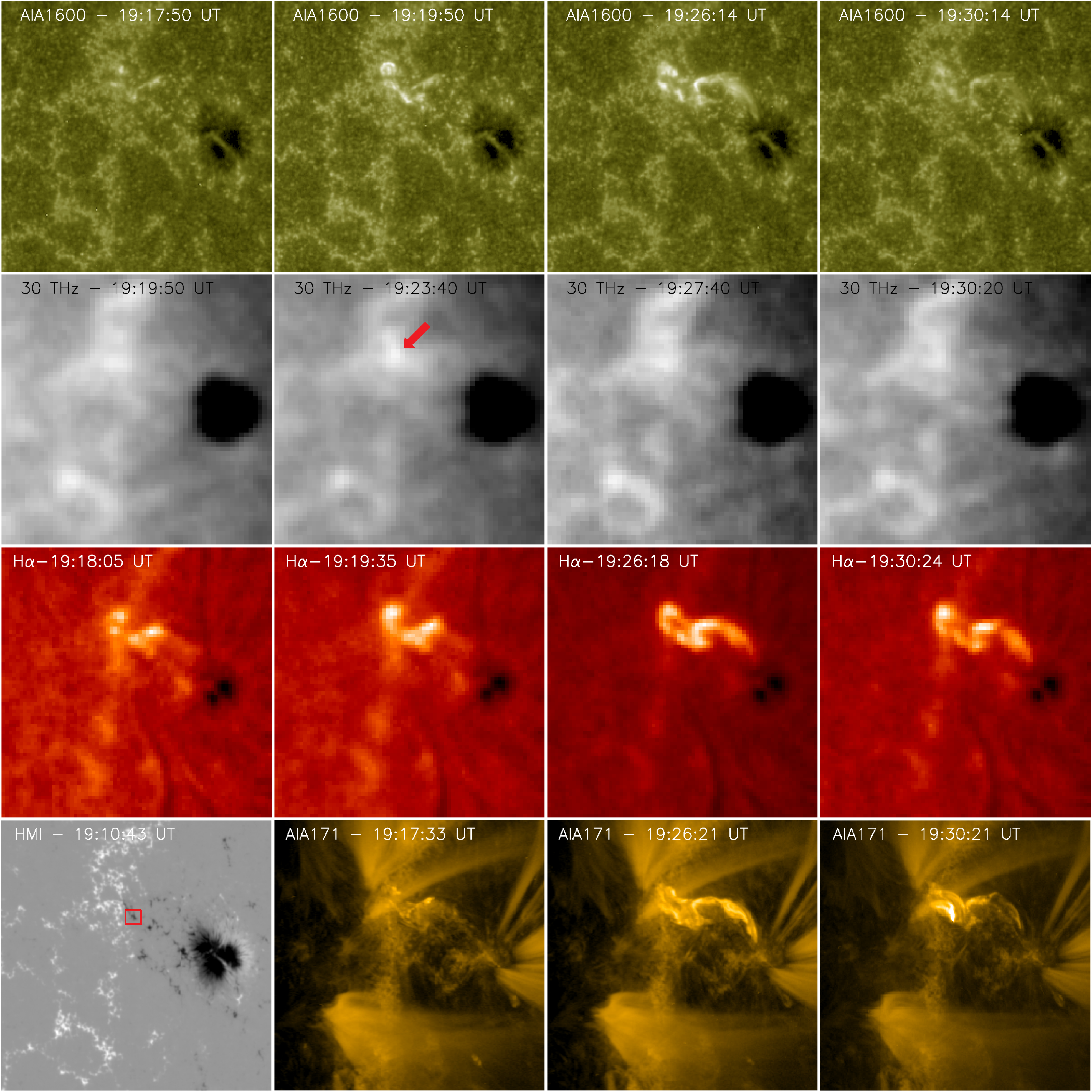}
\caption{Sequence of images of AR 12741 during the flare. The different rows show, from top to bottom, images of AIA1600, mid-IR, H$\alpha$, and a combination of HMI and AIA171 maps, between 19:15 to 19:30 UT, covering the whole event. The locations of flare kernels at flare onset are clear in the second panel of AIA1600, while later some loop structures appear bright, and by its end a small kernel is seen on the large negative polarity (see also the movie AIA1600.mp4 available online). 
In the second panel of the second row, the red arrow indicates the location of the flaring region in the 30~THz data, which is comparable to the location with AIA1600. These 30 THz features can be better seen in panel b) of Fig.~\ref{fig:kernels30THz}.
In the third row, the kernels are well observed in H$\alpha$ and they correspond well to the ones of AIA1600. At upper coronal heights, the flare evolution can be followed in the three last panels, last row, in AIA171 (see also the movie AIA171.mp4 available online). A clear small loop structure is seen connecting the kernels seen in AIA1600 and a larger elongated structure is seen to the west towards the spot. The first panel in this row depicts the HMI magnetic field as a reference for the location and polarity sign of the different flare features. The red square in this panel encompasses the region where we compute the negative flux variation shown in \fig{flux-evol}. All panels cover the same field of view; their sizes and the convention for the magnetic field maps are the same as in \fig{mag-arevol}.}
\label{fig:mapsAR}
\end{figure*}

\begin{figure}[ht!]
\resizebox{\hsize}{!}{\includegraphics{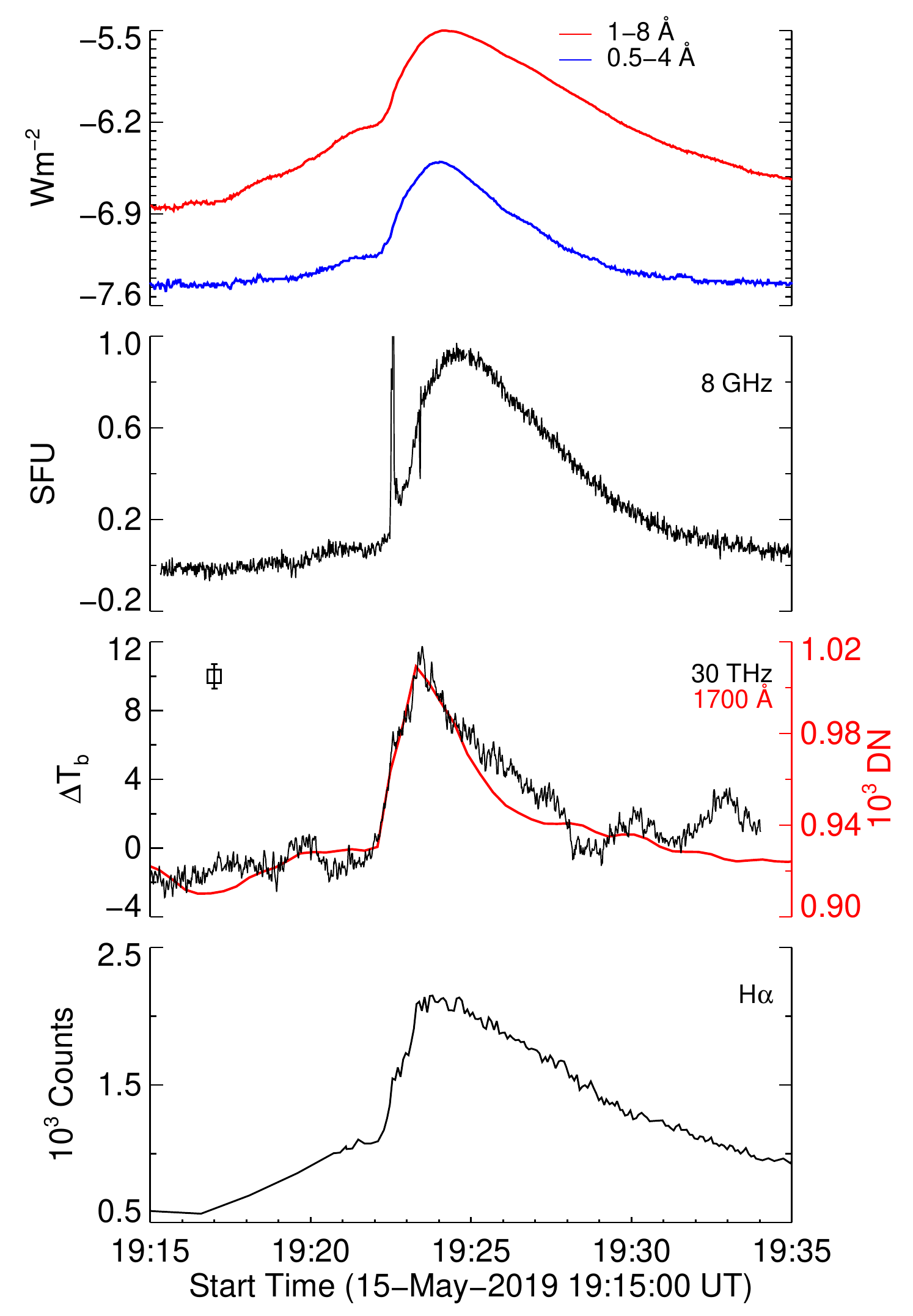}}
\caption{Temporal evolution at selected wavelengths. From top to bottom: GOES soft X-rays 1-8 and 0.5-4~\AA, EOVSA 8~GHz, AR30T mid-IR (30 THz), AIA 1700~\AA~and HASTA H$\alpha$. The error bar shown at the top left of the 30~THz plot corresponds to 1$\sigma$ uncertainty in $\Delta T_\mathrm{b}$. The sharp peak visible in the 8~GHz curve during the impulsive phase of the flare extends up to $\sim$3~SFU. SFU = solar flux unit. }
\label{fig:curves}
\end{figure}

The UV evolution of the flare was studied using the AIA1600 and AIA1700 data, where both filters show a similar behavior. 
The first row of \fig{mapsAR} corresponds to the flaring area as observed by AIA1600 at different times (see also the movie AIA1600.mp4). At around 19:17 UT, when a slight increase in GOES (see \fig{curves}) 1--8~\AA~curve is seen, four flare brightenings are observed; two are located on the positive field region, while the one to the south-west lies on a small negative polarity and an extension towards the west appears to the north of the latter kernel. As the flare evolves these brightenings extend and by 19:26 UT, 2 minutes after flare maximum in GOES, a faint elongated structure extends towards the main negative polarity. When this evolution is seen in the movie AIA1600.mp4, one can observe plasma flowing along this feature. By 19:30 UT, a small kernel appears to the north of the main negative polarity. This latter kernel is probably the counterpart of the northern flare kernel to the east seen earlier at 19:17 UT (\fig{mapsAR} top right). 

Using AIA images corrected by differential rotation, we integrated the signal in a 2~arcmin square box centered in the flaring region. The box size ensures that all the area associated with the flare is included on it. The resulting curve for AIA1700 is shown in \fig{curves}, together with that of the mid-IR (see Sect. \ref{sec:midIR-evol}) for a better comparison. The UV and mid-IR lightcurves show a remarkable similarity, suggesting that these emissions originate from the same region. 

The behavior of the event in H$\alpha$ is also similar to that observed in AIA1600. The evolution of the event as registered by HASTA is shown in the third row of \fig{mapsAR}. At 19:18 UT initial bright kernels are observed in the region, with spatial correspondence with the multiple brightenings observed in AIA1600. At 19:19 UT the multiple bright kernels are clearly visible increasing their brightness as the flare evolves. The flow of plasma described above for AIA1600 is also visible in this wavelength. The last panel in \fig{curves} shows the temporal evolution profile of the flare in H$\alpha$. As seen in the GOES 1-8~\AA~profile, at 19:17 UT a slow rise of chromospheric emission is detected in this filter. At 19:20:40 UT HASTA starts the flare observing mode. To extract the temporal evolution of the H$\alpha$ intensity and correct by the atmospheric opacity absorption,  we use the same technique applied by \citet{Kaufmann13}: for every frame we consider the flare intensity as the arithmetic mean inside an inner circle around the flaring region and subtract the background defined as the signal integrated over an outer ring. 

The last row in \fig{mapsAR} shows the evolution in AIA171 in the last three panels, see also the movie AIA171.mp4. Before flare maximum, extended bright structures appear to be present in between AIA1600 initial brightenings at 19:17 UT. AIA171 images show the presence of two loop-like structures: a small one (better seen at around 19:30 UT) that seems to connect the two initial AIA1600 kernels to the south and a longer one through which plasma is seen to flow towards the west. This long bright loop-like structure seems to be anchored in the northern AIA1600 kernel on the positive field region at one end and on the AIA1600 small kernel located to the north of the main negative polarity to the west. The first panel is a magnetic map included as reference and the red square surrounds the region where we have computed the negative magnetic flux shown in \fig{flux-evol}.

\subsection{Mid-IR evolution}
\label{sec:midIR-evol} 

Observations at 30~THz are shown in the second row of \fig{mapsAR}. A region 20\arcsec\ north and 50\arcsec\ east from the sunspot, marked by a red arrow in the second panel, changes its brightness between frames. 

\begin{figure}[ht!]
\begin{center}
\begin{tabular}{c c}
\includegraphics[height=0.23\textwidth]{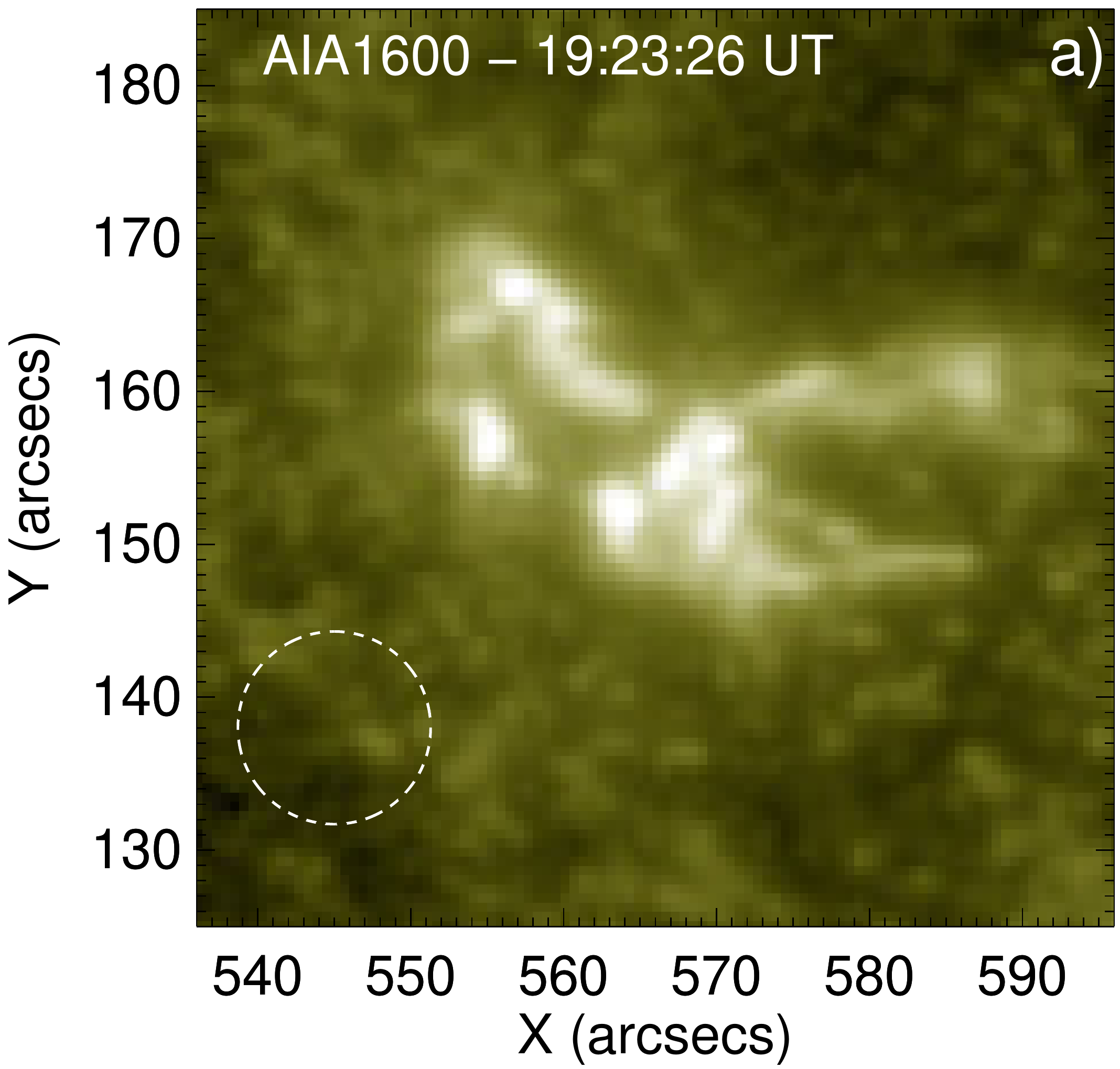}\hspace{-0.45cm} & \includegraphics[height=0.23\textwidth]{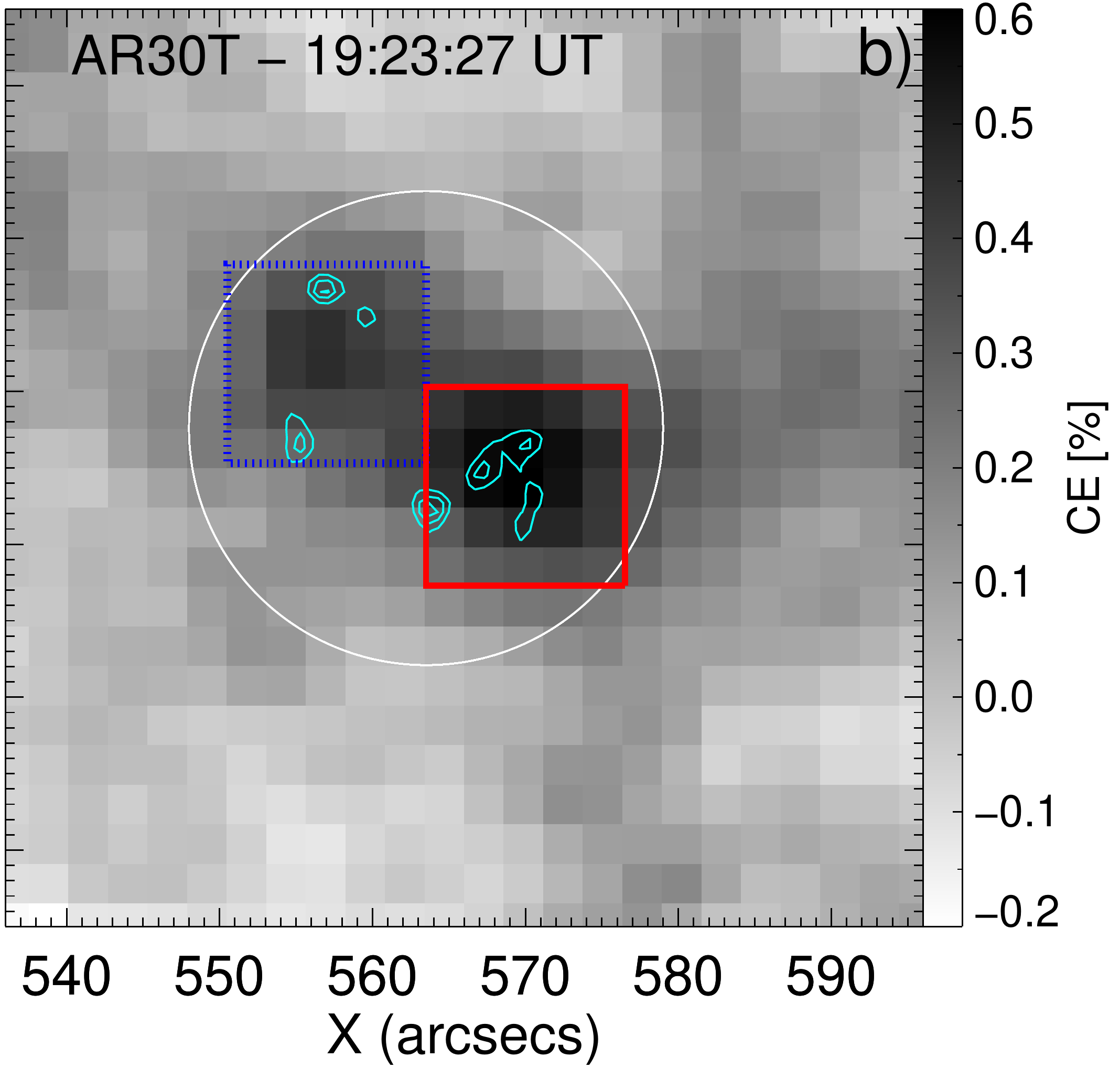}
\end{tabular}
\includegraphics[width=0.4\textwidth]{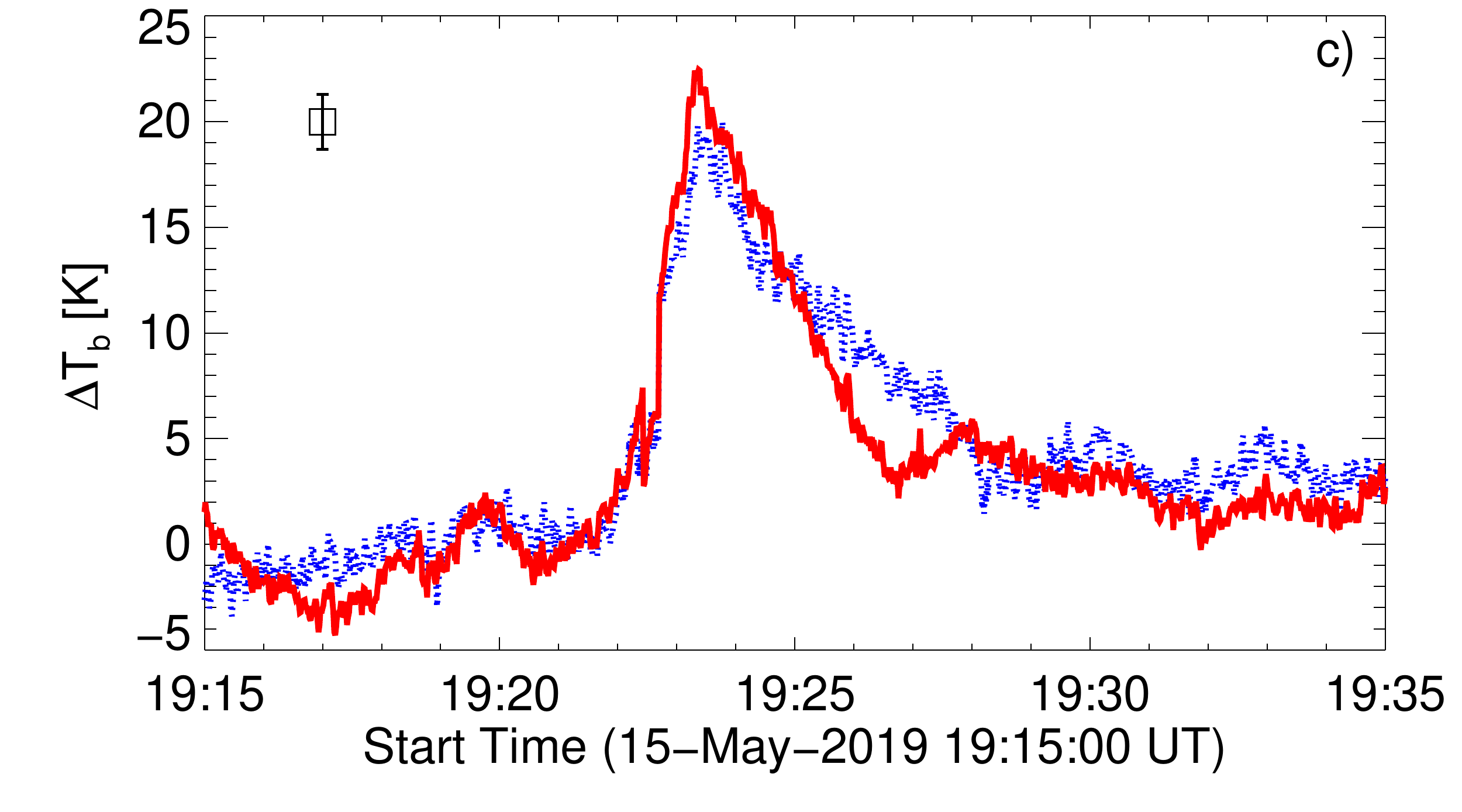}
\caption{Close view of the mid-IR and UV sources. Panel a) shows the flaring region observed by AIA1600. The dashed white circle indicates the size of the Airy disk of the AR30T. Panel b) displays the two bright kernels observed in 30~THz after pre-event subtraction with colors inverted (black is more intense). The dotted-blue and red boxes enclose the northern and southern kernels respectively. The contour levels are 50, 70, and 90\% of the flare maximum in AIA1600. The white circle indicates the area used to estimate the mid-IR temporal evolution displayed in \fig{curves}. The excess brightness temperature averaged for the boxes shown in the upper panel (with the respective blue and red color) is shown in panel c). The error bar indicates 1$\sigma$ uncertainty in $\Delta T_\mathrm{b}$.}
\label{fig:kernels30THz}
\end{center}
\end{figure}

To obtain the temporal evolution of the flaring region, we use the same technique described for H$\alpha$ data (see Sect.~\ref{sec:chromos_EUV}). To calibrate the pixel intensities above the pre-flare level into excess brightness temperature $\Delta T_\mathrm{b}$, we use the quiet Sun as a reference, assuming T$_\sun=5\,000$~K at 30~THz \citep{Turon70}. Then, for every frame, the quiet Sun raw pixel temperature with respect to the sky raw temperature is measured and a scale is obtained to convert the raw to brightness temperature. The pre-flare level is defined as the brightness temperature of the flaring area averaged between 19:15 UT and 19:20 UT, with the standard deviation (1$\sigma$) as its uncertainty. This pre-flare level is subtracted for every frame. Figure \ref{fig:curves} shows the average excess brightness temperature over the region delimited by the white circle shown in the panel b) of \fig{kernels30THz}, which encloses all the mid-IR flaring area.

Panel b) in \fig{kernels30THz} shows the inverted-intensity (black is more intense) image of the flaring region after the subtraction of a background ($\bar{{I}}_\mathrm{pre}$) image built by averaging 60 successive pre-event images. The two bright kernels associated  with the event become evident only by using difference images. The bright sources are separated by $\approx$18\arcsec, probably associated with chromospheric footpoints of magnetic loops \citep{Pennetal:2016}. The sides of the boxes correspond to the size of the telescope Airy disk. The location of these kernels can be easily compared to the location of the brightest regions in AIA1600 in panel a) to its left. The color bar shows the contrast excess (CE = $({I}_\mathrm{fl} - \bar{{I}}_\mathrm{pre}$) / $\bar{{I}}_\mathrm{pre}$) for the flaring region. When we restrict the area over which we compute the excess brightness temperature evolution to the boxes around the kernels where most of the mid-IR radiation is emitted,  we get the curves shown in panel c) of \fig{kernels30THz}.  As it is evident, both kernels share a similar temporal evolution, as well as temperature excess values that are higher than the one obtained before for a bigger box (see \fig{curves}) . 

In the mid-IR regime, the point spread function (PSF) of small telescopes is determined by the diffraction pattern \citep{Turon70,Trottet15,Miteva16,GimenezdeCastroetal:2018}, which limits the instrument spatial resolution. Given that the kernels in \fig{kernels30THz} are separated by a distance comparable to the PSF, it is reasonable to expect a small amount of contamination between them when trying to separate the contribution of each individual kernel. 
Therefore, we average the excess brightness temperatures over these two boxes together to obtain the time profile of the flaring area shown in \fig{curves}. The rise phase of the flare is in excellent agreement in the mid-IR, UV, and H$\alpha$ spectral bands, while the decay phase agrees very well with that of the UV band. The maximum intensity of the flare at 30~THz is registered at 19:23:27 UT and coincides with that of AIA1700 within its $\pm~12$~s temporal resolution. 

\subsection{Microwaves: Absence of a signature from accelerated electrons}
\label{sec:MW}

\begin{figure}
\includegraphics[width=0.5\textwidth]{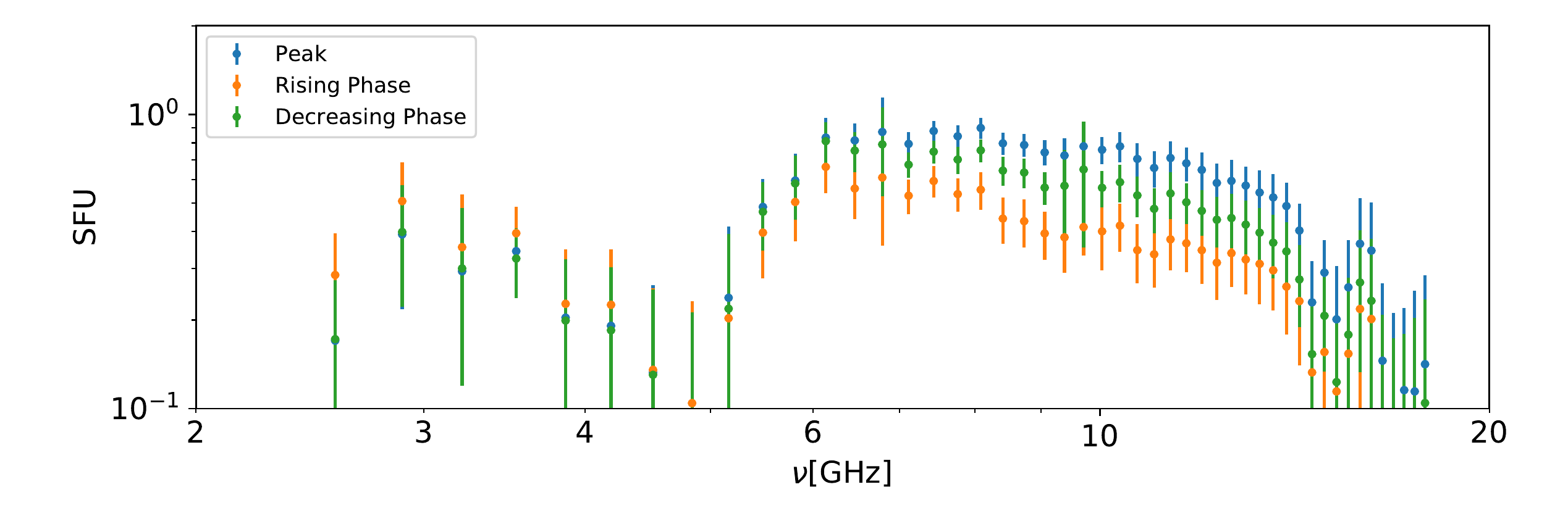}
\caption{Microwave spectra obtained at different phases of the flare. Each spectrum is a 90~s time integration corresponding to the rising (orange dots), peak (blue dots), and decreasing phase (green dots). The vertical bars represent the standard deviation of the mean.}
\label{fig:MWSpec}
\end{figure}

EOVSA spectral data at 8 GHz were binned in 50 spectral windows of 0.325~GHz bandwidth each, measured serially at 20~ms sample time, covering the 50 bands in 1~s. The light curve at 8~GHz, around the spectral maximum, shows a very low flux density, below 1~SFU, and a time evolution very similar to that of GOES soft X-rays, with the peak delayed around 60~s relative to GOES and 30~THz peaks (\fig{curves}). This gradual evolution is an unusual behavior for the MW lightcurves, which often show an impulsive time profile \citep{White11}. At the beginning of the event a short ($\simeq 3$~s) and a relatively strong spike\footnote{\textit{relative} to the maximum of this flare; however it is still weak in an absolute sense, $< 3$~SFU.} can be seen. We analyzed the spike spectrum (not shown here) and found it to be very narrowband. This feature is similar to the one analyzed by \citet{GimenezdeCastroetal:2006}, its investigation, however, is beyond the scope of this paper. 

In \fig{MWSpec} we show time integrated spectra taken at different phases of the event: rising, peak, and decreasing phase. There are no significant differences between them, besides the flux density, and all of them are rather flat (within the instrumental uncertainties) at frequencies above about 5~GHz, lending support to an optically thin thermal emission \citep{Dulk85}. The decreasing flux above 15~GHz is more uncertain due to the rising solar background, thus reducing the detection sensitivity at such low intensities, and remains consistent with a flat spectrum. The absence of a clear signature of gyrosynchrotron emission from high energy electrons \citep[\eg~Fig. 2 in][]{Dulk85} in the observed spectra in this flare, and the gradual evolution of the MW light curves (which are well associated with the SXR curves) suggest that the MW emission is produced mainly by free-free radiation. From that, it follows that this flare did not produce a significant number of accelerated electrons and, therefore, such electrons cannot be responsible for the main energy transport.

\section{Discussion}
\label{sec:Discussion}

This section is divided into a discussion of this study's three main points. In the first, we discuss how the evolution of the photospheric magnetic field assigned a location to the reconnection site. Then we make a better estimate of the temperature excess brightness at mid-IR. Finally,  we discuss thermal conduction as the main driving mechanism for the mid-IR emission.  

\subsection{Proposed flare scenario}
\label{sec:scenario}

AR 12741 was located at N06W45 on the flare day; therefore, projection effects impacted the line of sight magnetic field measurements (see the false positive polarity that appeared to the west of the main negative spot already at the beginning of 15 May in the movie HMI-8-17May.mp4).

\begin{figure}
\includegraphics[width=0.5\textwidth]{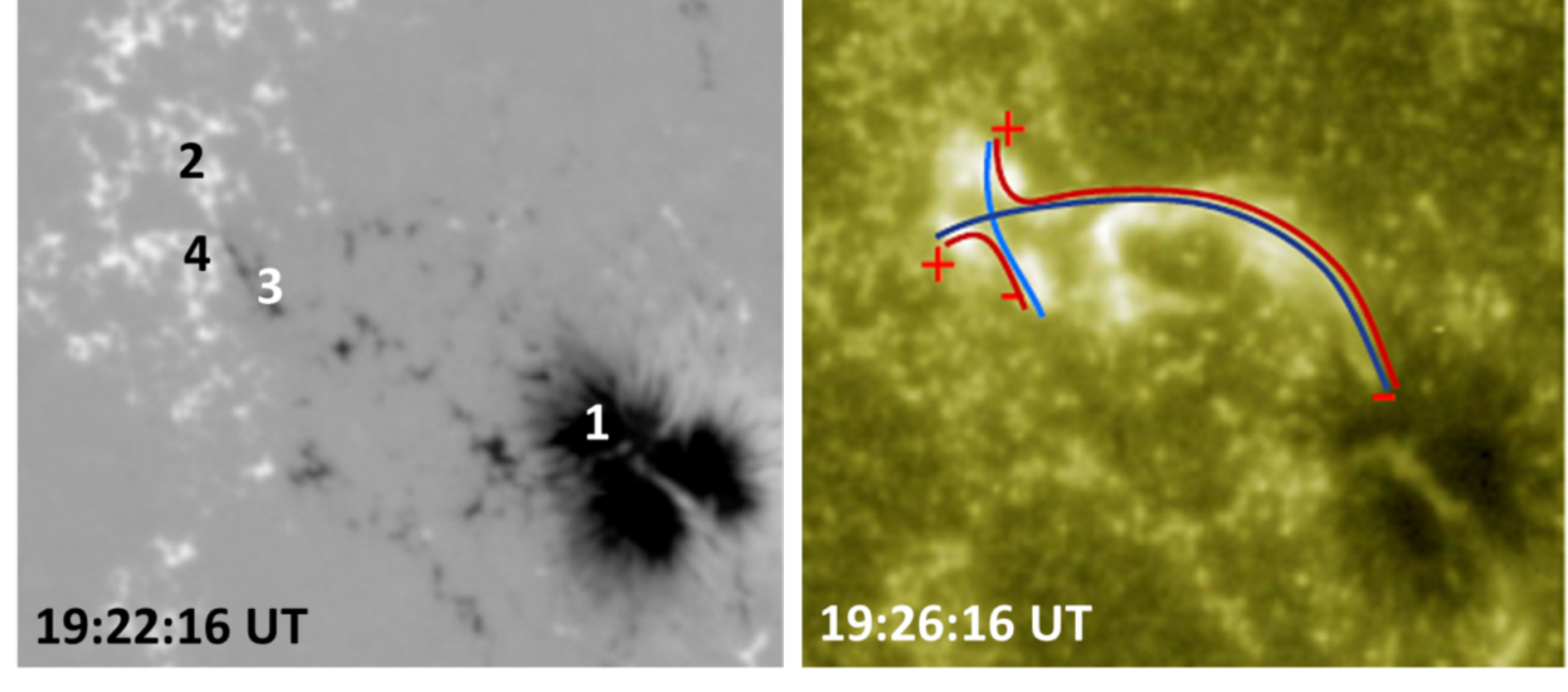}
\caption{Close up view of the photospheric magnetic field configuration. Left panel: SDO/HMI line-of-sight magnetogram of AR 12741 on 15 May 2019 at around flare maximum. The numbers indicate the polarities where the loops associated with the energy release during the flare could be anchored. Right panel: AIA1600 image including a sketch of the connectivities related to the loops involved in the flare. Blue and light blue continuous lines indicate loops before reconnection, while red lines would correspond to reconnected loops. Plus and minus signs mark the polarity sign of the footpoints. Both panels cover the same field of view; their sizes are 94\arcsec in the horizontal and 87\arcsec in the vertical directions. The convention for the magnetic field map is the same as in \fig{mag-arevol}.}
\label{fig:topology}
\end{figure}

\Fig{topology} (left panel) shows an HMI magnetogram at around flare maximum. We have numbered the polarities that, according to the evolution of the event described in Sect. \ref{sec:chromos_EUV} and shown in \fig{mapsAR}, are associated with the footpoints of the small bright loop seen in the AIA171 image at 19:26:21 UT (polarities 3-4) and the elongated one connecting to the main negative spot (polarities 1-2). The latter is the one along which we observe plasma flowing starting at $\approx$19:22 UT, as observed in movies AIA171.mp4 and AIA1600.mp4. These two loops could be the result of the magnetic reconnection process at the origin of the C2.0 flare. In the right panel, we traced these two loops with continuous red lines. Considering that these loops result from reconnection in a quadrupolar magnetic configuration \citep[see an example in][]{Mandrini14}, the two continuous light blue (connecting polarities 2-3) and blue (connecting polarities 4-1) lines correspond to the pre-reconnected loops. Because of the aforementioned projection effect problems, we did not attempt to model the AR magnetic field and show a sketch in the right panel of \fig{topology}. Furthermore, reconnection between a higher plasma pressure loop like the one connecting 2-3 with the one connecting 4-1, could drive the injection of material into the longer loop between polarities 2 and 1, as we have observed. For more details, we refer to  a similar process proposed in \citet{Baker09} and \citet{Mandrini15}, and illustrated in Fig. 5 of \citet{vanDriel12}, to explain the origin of plasma upflows observed at the border of active regions in the EUV.

We computed the magnetic flux variation in the region corresponding to the small negative polarity 3 (see \fig{flux-evol} and the inset in the same panel). From this analysis, a clear decrease of the negative magnetic flux is observed from at least two hours prior to the flare. We interpret this result as mainly due to flux cancellation with the nearby positive polarity.
The negative flux decrease rate increased from about 2~minutes before the impulsive phase. This decreasing steep slope stayed nearly constant during the flare excess emission in mid-IR. When the flare excess ceased, the flux decrease ceased almost simultaneously. 
Magnetic flux changes in line-of-sight magnetograms, with a tendency toward flux decreases in a step-wise manner, have been observed in association with flares higher than GOES M5 class by \citet{Burtseva13} (see also references in that article) during their impulsive phase. These results were interpreted as consistent with a model of collapsing loop structures \citep[see][and references therein]{Hudson08}. Our flare is a less energetic one compared to those studied in that article; however, we clearly see a step-wise flux change as found in the most energetic events (see \fig{flux-evol}).

\subsection{Estimating the maximum IR brightness temperature}\label{sec:filling_factor}

It is expected that the 30~THz excess brightness temperature will be underestimated because of the telescope spatial resolution. If 30~THz originates in the same place where AIA 1600~\AA\ is emitted, as \fig{kernels30THz} suggests, we can estimate a filling factor for mid-IR.  We recall that the camera has a diffraction limit greater than the pixel size; therefore, we should take the area defined by a solid angle with a diameter equal to the diffraction limit, which corresponds to 50\% of the full width at half maximum (FWHM) of the PSF of the telescope. The diffraction limit of 12.6\arcsec\ corresponds to an area equal to $A_\mathrm{PSF} = 6.57\times 10^{17}\ \mathrm{cm}^2$, and from \fig{kernels30THz}a), we see that the whole UV emitting area extends over two telescope PSFs. Using the 50\% level contours of \fig{kernels30THz} we obtain a 1600~\AA\ emitting area equal to $A_\mathrm{flare} = 1.43 \times 10^{17}\ \mathrm{cm}^2$. 
The filling factor $f\!f$ is thus obtained  as: 
$$f\!f = \frac{A_\mathrm{flare}}{2 A_\mathrm{PSF}} = 0.11\ .$$
Therefore we can derive a deconvolved\textit{} excess brightness temperature: 
$$\Delta T^*_\mathrm{b} = \frac{\Delta T_b}{f\!f} = \frac{20}{0.11} = 182\ \mathrm{K} \ . $$
This deconvolved temperature is still lower than the values obtained by \cite{Simoesetal:2017} via radiative hydro-dynamic simulations, which used non-thermal electrons to drive the chromospheric heating. In those simulations, the lowest brightness temperature peak excess \citep[about 300K; see Figure 6c in][]{Simoesetal:2017} was found in the case for an electron beam with a low energy cut-off of 20 keV and spectral index of 8. This follows a trend identified in that study that the peak brightness temperature correlates well with the amount of energy deposited deeper in the chromosphere, where more hydrogen can be ionized and thus increases the IR free-free emission. In our case, with such low peak brightness temperature, we speculate that only the upper layers of the chromosphere were efficiently heated and ionized. We can infer that this is consistent with either a very steep distribution of non-thermal electrons or a thermal conducting front. Given the lack of observational evidence for accelerated electrons (see Section \ref{sec:MW}), by elimination, we conclude that the chromospheric heating was driven by a conduction front in this event.
   
For the flaring region in the mid-IR we can estimate the flux density, $S_\nu$, using the Rayleigh-Jeans approximation to a black body:
\begin{equation}
    S_\nu = \frac{2k_\mathrm{B} \nu^2}{c^2} \ \Delta T^*_\mathrm{b}\  \Delta \Omega = 
    322 \ \mathrm{SFU} , 
\label{Rayleigh-Jeans}
\end{equation}

where $k_\mathrm{B}$ is the Boltzmann constant, $c$ is the speed of light in vacuum, $\nu$ = 30~THz, and $\Delta \Omega$ is the solid angle subtended by the flaring region corresponding to the 1600~\AA\ area. We note that the flux density is independent of the filling factor, $f\!f,$ since when $\Delta T^*_b \propto f\!f^{-1}$ -- the solid angle is $\Delta \Omega \propto f\!f$ and then its effect is canceled out. 

\subsection{Evaluation of thermal conduction as the main energy transport mechanism}
\label{sec:Conduction}

Since there is no evidence of an intense flux of accelerated particles precipitating in the chromosphere, we suggest that the energy that drives the chromospheric response (\ie~the IR, H$\alpha,$ and UV emission) is primarily the result of thermal conduction. We follow \cite{Battagliaetal:2009} to evaluate the energy flux transported by conduction from a hot coronal source located approximately midway between the loop footpoints:
\begin{equation}
    F_\mathrm{cond} \simeq 10^{-6} \varrho({\cal R}) \frac{T_\mathrm{cs}^{7/2}}{L_\mathrm{loop}} \ [\mathrm{erg\ cm^{-2}\ s^{-1}}] \ , 
\label{cond}
\end{equation}
where $T_\mathrm{cs}$ is the coronal source temperature, $L_\mathrm{loop}$ is half loop length and $\varrho({\cal R})$ is a reduction factor to get an effective heat flux. 
   $${\cal R} = \lambda_\mathrm{emf}/L_\mathrm{th}, $$ 
with $L_\mathrm{th}$ is the temperature scale length, and the electron mean free path is defined as:
   $$\lambda_\mathrm{emf} = 5.21 \times 10^3 \frac{T^2}{n_\mathrm{e}}. $$
Finally,
   $$\varrho({\cal R}) =  1.01 \cdot e^{-0.05 (\ln{\cal R} + 6.63)^2} < 1 \ . $$
As in \cite{Battagliaetal:2009}, we assume that $L_\mathrm{th} \simeq L_\mathrm{loop}$. 

Following our discussion in Sect.~\ref{sec:scenario} and considering, as a first-order approach, that the small bright loop, connecting polarities 4 and 3, has a semicircular shape, we can estimate its length as 30~Mm. Its half-length value can be used in Eq.~\ref{cond} to estimate the energy flux transported by conduction, \ie~$1.5 \times 10^9\ \mathrm{cm}$. We chose this short loop because it is the closest to the two 30~THz kernels whose origin we aim to explain. We speculate that a similar amount of energy could have been input in the longer loop connecting 2-1; however, we do not  observe mid-IR (30~THz) emission associated with it. Therefore, we consider our choice of the half length of the shorter loop to be good for an order of magnitude estimation of the conductive energy flux. Assuming emission from an isothermal plasma we use GOES to infer the coronal parameters during the flare and obtain an emission measure $EM = 1\times10^{48}\ \mathrm{cm}^{-3}$ and a $T_\mathrm{cs} = 12\ \mathrm{MK}$. Considering the product of the flare emitting area in AIA1600 and the altitude of the semicircular loop, we estimate a flaring volume of $1.36\times10^{26}\ \mathrm{cm}^{3}$, getting a density $n_\mathrm{e} = 8.6\times 10^{10}\ \mathrm{cm}^{-3}$, then: 
$$ {\cal R} = \frac{5.21 \times 10^3 \ (12\times 10^6)^2 /8.6\times 10^{10}}{1.5 \times 10^9} = 0.0058 \ , 
$$ yielding a $\varrho({\cal R}) = 0.90$ and an energy flux: 
$$ F_\mathrm{cond} \simeq 0.90 \times 10^{-6} 
\frac{(12\times 10^6)^{7/2}}{1.5\times 10^9} = 3.61\times 10^9  \ [\mathrm{erg\ cm^{-2}\ s^{-1}}] \ . $$

A flare duration of $\Delta t = 326$~s is estimated from the mid-IR temporal profile (see \fig{curves}), as the time interval while $\Delta T_\mathrm{b}$ is above the pre-flare level plus $5\sigma$. If we additionally take the UV emitting area considered in Sect.~\ref{sec:filling_factor}, 
we estimate a total conductive energy: 
$$E_\mathrm{cond} = 1.43 \times 10^{17} \cdot 326 \cdot 3.61\times 10^9 \approx 1.68\times 10^{29}\ \mathrm{erg}.$$

We can compare this number with the estimate of the energy emitted in the mid-IR range
considering that the camera has a band pass $\Delta \nu=17$~THz and $R=1$~astronomical unit (AU) and $t_f - t_i = \Delta t$:
$$E_\mathrm{30T}= \Delta \nu \times 2\pi \ R^2 \times \int_{t_i}^{t_f} F_\mathrm{30T}(t) \,dt \sim 10^{26} \ \mathrm{erg} \ll E_\mathrm{cond} \ .$$
Here, $E_\mathrm{cond}$ is of the same order of magnitude of the total radiated flare energy $E_\mathrm{rad} \approx 2.6 \times 10^{29}\ \mathrm{erg}$, as estimated from the Extreme ultraviolet Solar Photometer (ESP), part of the Extreme ultraviolet Variability Experiment \citep[EVE, ][]{Woods12}. The value for $E_\mathrm{rad}$ was estimated by taking the time-integrated, pre-flare-subtracted, irradiance measured by the ESP channels 16.64--21.50~nm, 22.28--28.78~nm, 27.16--33.80~nm, and 0.1--7.0~nm, and summed together. The total flux density was converted to energy irradiated at the Sun by multiplying by $2\pi \ R^2$.

Therefore, with $E_\mathrm{cond} \simeq E_\mathrm{rad}$ and the fact that the EOVSA MW spectrum shows no signature of accelerated electrons, we conclude that thermal conduction is likely to be the main energy transport mechanism in this event. Moreover, the estimated deconvolved brightness temperature 182~K puts this event below the values obtained by the simulations by \citet{Simoesetal:2017}, where the IR emission is the result of the solar atmosphere being heated by accelerated electrons. Those simulations show that softer electron beams produce weaker IR emission than harder electron beams. This is because the softer beam cannot penetrate deep into the chromosphere and cause H ionisation where the  density is higher. Extending this idea, a thermal conduction front would mostly heat the top of the chromosphere, producing comparatively lower numbers of free electrons, and thus, weaker IR emission than in the case of a beam-driven flare.

\section{Conclusions}
\label{sec:Conclusions}

In this work, we present an analysis of a GOES C2.0 solar flare observed at mid-IR by the ground-based telescope AR30T. Specifically, SOL20190515 is the weakest event ever reported at mid-IR in terms of its flux density (about 300~SFU) and brightness temperature (about 20~K before deconvolution). We summarize our main findings and conclusions as follows: 

\begin{itemize}

\item Previous works \citep[\eg][]{Kaufmann13, GimenezdeCastroetal:2018} have considered mid-IR emission as a good proxy of white-light emission during solar flares. However, for this event, we found no evidence of white-light emission in HMI pseudo-continuum data, although it shows a clear mid-IR emission. Telescopes observing in the visible band with higher sensitivity, such as the Daniel K. Inouye Solar Telescope \citep[DKIST,][]{Rast21} or the European Solar Telescope \citep[EST,][]{EST19}, should be able to detect the flare excess (or lack thereof) in white-light for this kind of weak events \citep{Jess08}. 

\item The similar behavior between the time profiles of 30 THz and AIA1700, first reported in \citet{Miteva16} for an impulsive GOES M2 class flare, is confirmed in our work for the thermal-driven event analyzed here. Both emissions have a chromospheric origin. However, the similarity is not evident since the mechanisms responsible for them are different. The 30 THz radiation during flares is due to thermal free-free bremsstrahlung \citep{Trottet15, Simoesetal:2017}, while the flare excess emission for AIA1600 and AIA1700 is dominated by the contribution of several chromospheric spectral lines, with formation temperatures ranging from $10^4$ to $10^5$~K \citep{Simoesetal:2019A}. 

\item The magnetic field evolution of the AR, where the flare occurs, is marked by its decay phase (presence of light bridges and MMFs related to its main preceding spot). This evolution combined with the observations in all analyzed AIA bands, led us to propose a flare scenario in which the event originates at low coronal heights and to infer the geometrical parameters that we later use to compute the main flare energy input. 

\item The flat spectra obtained at microwaves, as well as the similar temporal profiles observed in microwaves and soft X-rays, led us to conclude that there is no clear evidence for the presence of a significant number of non-thermal electrons. Additionally, the "deconvolved" brightness temperature excess of 182~K found in mid-IR is still lower than the minimum value expected from the injection of non-thermal particles according to the hydrodynamics simulations by \citet{Simoesetal:2017}. Although we cannot discard 
other transport mechanisms, based in our findings, we consider thermal conduction as the dominant mechanism responsible for the energy transport in this flare. This conclusion is supported by our estimates of the total thermal conduction energy, for which our proposed flare scenario provides the required geometrical parameters, and the total radiated energy that represent the total energy input and output of the flare. Both values are within the same order of magnitude, when calculated under reasonable assumptions.

\item The AR30T telescope has been proven to have a high enough sensitivity to detect the mid-IR emission from C class flares, given it is mainly limited by its spatial resolving power. The strong association between (ground-based) mid-IR and UV observations, as well as the high cadence of data acquisition, makes the AR30T an excellent instrument that ought to be considered as a much less expensive proxy of space-based chromospheric continuum UV observations. 

\item With the rise of solar activity in the next months and years, we expect to increase our capabilities to examine the mid-IR emission in solar flares by complementing the observations by the AR30T with the data from the future High Altitude THz Solar Photometer \citep[HATS,][]{GimenezdeCastro20}, which is currently at the final stages of installation. 

\end{itemize}

This work reinforces the key role of the mid-IR diagnostics to describe the Chromosphere dynamics during flares, which is more relevant for weak flares as the present study has shown. Interest in weak flares has increased in present days because of the new instrumentation on board of the solar orbiting probes with greater sensitivity and spatial resolution. However, mid-IR instruments work with high sensitivity even when installed on the ground. In future works we expect to have more multi-wavelength flare analysis combining X-rays and EUV data obtained by the solar orbiting probes and by our mid-IR cameras.

\begin{acknowledgements}

This research was partially supported by FAPESP grant 2013/24155-3 and CAPES 88887.310385/2018-00. FML is a CAPES fellow. PJAS acknowledges support from the Fundo de Pesquisa Mackenzie (MackPesquisa) and CNPq (contract 307612/2019-8). CM and GDC acknowledge grants PICT 2016-0221 (ANPCyT) and UBACyT 20020170100611BA. GDC is a member of the Carrera del Investigador Cient\'\i fico of the Consejo Nacional de Investigaciones Cient\'\i ficas y T\'ecnicas (CONICET). CHM is a CONICET researcher. CGGC is a corresponding researcher of CONICET and fellow of the CNPq (PQ level 2, grant 307722/2019-8). DG and EOVSA are supported by NSF grant AST-1910354 and NASA grants 80NSSC18K1128 and 80NSSC20K0627 to New Jersey Institute of Technology. We recognize the collaborative and open nature of knowledge creation and dissemination, under the control of the academic community as expressed by Camille No\^us at \href{http://www.cogitamus.fr/indexen.html}{http://www.cogitamus.fr/indexen.html}. 

\end{acknowledgements}
\bibliographystyle{aa} 
\bibliography{SOL20190515} 

\begin{thebibliography}{42}
\expandafter\ifx\csname natexlab\endcsname\relax\def\natexlab#1{#1}\fi

\bibitem[{{Allred} {et~al.}(2015){Allred}, {Kowalski}, \&
  {Carlsson}}]{Allred15}
{Allred}, J.~C., {Kowalski}, A.~F., \& {Carlsson}, M. 2015, \apj, 809, 104

\bibitem[{{Bagal{\'a}} {et~al.}(1999){Bagal{\'a}}, {Bauer}, {Fern{\'a}ndez
  Borda}, {Francile}, {Haerendel}, {Rieger}, \& {Rovira}}]{Bagala99}
{Bagal{\'a}}, L.~G., {Bauer}, O.~H., {Fern{\'a}ndez Borda}, R., {et~al.} 1999,
  in ESA Special Publication, Vol.~9, Magnetic Fields and Solar Processes, ed.
  A.~{Wilson} \& {et al.}, 469

\bibitem[{{Baker} {et~al.}(2009){Baker}, {van Driel-Gesztelyi}, {Mandrini},
  {D{\'e}moulin}, \& {Murray}}]{Baker09}
{Baker}, D., {van Driel-Gesztelyi}, L., {Mandrini}, C.~H., {D{\'e}moulin}, P.,
  \& {Murray}, M.~J. 2009, \apj, 705, 926

\bibitem[{{Battaglia} {et~al.}(2009){Battaglia}, {Fletcher}, \&
  {Benz}}]{Battagliaetal:2009}
{Battaglia}, M., {Fletcher}, L., \& {Benz}, A.~O. 2009, \aap, 498, 891

\bibitem[{{Burtseva} \& {Petrie}(2013)}]{Burtseva13}
{Burtseva}, O. \& {Petrie}, G. 2013, \solphys, 283, 429

\bibitem[{{Dulk}(1985)}]{Dulk85}
{Dulk}, G.~A. 1985, \araa, 23, 169

\bibitem[{{Fernandez Borda} {et~al.}(2002){Fernandez Borda}, {Mininni}, {Mand
  rini}, {G{\'o}mez}, {Bauer}, \& {Rovira}}]{FernandezBorda02}
{Fernandez Borda}, R.~A., {Mininni}, P.~D., {Mand rini}, C.~H., {et~al.} 2002,
  \solphys, 206, 347

\bibitem[{{Gary} {et~al.}(2018){Gary}, {Chen}, {Dennis}, {Fleishman},
  {Hurford}, {Krucker}, {McTiernan}, {Nita}, {Shih}, {White}, \&
  {Yu}}]{Garyetal:2018}
{Gary}, D.~E., {Chen}, B., {Dennis}, B.~R., {et~al.} 2018, \apj, 863, 83

\bibitem[{{Gim{\'e}nez de Castro} {et~al.}(2006){Gim{\'e}nez de Castro},
  {Costa}, {Silva}, {Sim{\~o}es}, {Correia}, \&
  {Magun}}]{GimenezdeCastroetal:2006}
{Gim{\'e}nez de Castro}, C.~G., {Costa}, J.~E.~R., {Silva}, A.~V.~R., {et~al.}
  2006, \aap, 457, 693

\bibitem[{{Gim{\'e}nez de Castro} {et~al.}(2020){Gim{\'e}nez de Castro},
  {Raulin}, {Valio}, {Alaia}, {Alvarenga}, {Bortolucci}, {Fernandes},
  {Francile}, {Giorgetti}, {Kudaka}, {L{\'o}pez}, {Marcon}, {Marun}, \&
  {Zaquela}}]{GimenezdeCastro20}
{Gim{\'e}nez de Castro}, C.~G., {Raulin}, J.-P., {Valio}, A., {et~al.} 2020,
  \solphys, 295, 56

\bibitem[{{Gim{\'e}nez de Castro} {et~al.}(2018){Gim{\'e}nez de Castro},
  {Raulin}, {Valle Silva}, {Sim{\~o}es}, {Kudaka}, \&
  {Valio}}]{GimenezdeCastroetal:2018}
{Gim{\'e}nez de Castro}, C.~G., {Raulin}, J.~P., {Valle Silva}, J.~F., {et~al.}
  2018, Space Weather, 16, 1261

\bibitem[{{Harvey} \& {Harvey}(1973)}]{Harvey73}
{Harvey}, K. \& {Harvey}, J. 1973, \solphys, 28, 61

\bibitem[{{Heinzel} \& {Avrett}(2012)}]{Heinzel12}
{Heinzel}, P. \& {Avrett}, E.~H. 2012, \solphys, 277, 31

\bibitem[{{Hudson} {et~al.}(2008){Hudson}, {Fisher}, \& {Welsch}}]{Hudson08}
{Hudson}, H.~S., {Fisher}, G.~H., \& {Welsch}, B.~T. 2008, in Astronomical
  Society of the Pacific Conference Series, Vol. 383, Subsurface and
  Atmospheric Influences on Solar Activity, ed. R.~{Howe}, R.~W. {Komm}, K.~S.
  {Balasubramaniam}, \& G.~J.~D. {Petrie}, 221

\bibitem[{{Jess} {et~al.}(2008){Jess}, {Mathioudakis}, {Crockett}, \&
  {Keenan}}]{Jess08}
{Jess}, D.~B., {Mathioudakis}, M., {Crockett}, P.~J., \& {Keenan}, F.~P. 2008,
  \apjl, 688, L119

\bibitem[{{Kaufmann} {et~al.}(2013){Kaufmann}, {White}, {Freeland}, {Marcon},
  {Fernandes}, {Kudaka}, {de Souza}, {Aballay}, {Fernandez}, {Godoy}, {Marun},
  {Valio}, {Raulin}, \& {Gim{\'e}nez de Castro}}]{Kaufmann13}
{Kaufmann}, P., {White}, S.~M., {Freeland}, S.~L., {et~al.} 2013, \apj, 768,
  134

\bibitem[{{Kaufmann} {et~al.}(2015){Kaufmann}, {White}, {Marcon}, {Kudaka},
  {Cabezas}, {Cassiano}, {Francile}, {Fernandes}, {Hidalgo Ramirez}, {Luoni},
  {Marun}, {Pereyra}, \& {Souza}}]{Kaufmannetal:2015}
{Kaufmann}, P., {White}, S.~M., {Marcon}, R., {et~al.} 2015, Journal of
  Geophysical Research (Space Physics), 120, 4155

\bibitem[{{Lemen} {et~al.}(2012){Lemen}, {Title}, {Akin}, {Boerner}, {Chou},
  {Drake}, {Duncan}, {Edwards}, {Friedlaender}, {Heyman}, {Hurlburt}, {Katz},
  {Kushner}, {Levay}, {Lindgren}, {Mathur}, {McFeaters}, {Mitchell}, {Rehse},
  {Schrijver}, {Springer}, {Stern}, {Tarbell}, {Wuelser}, {Wolfson}, {Yanari},
  {Bookbinder}, {Cheimets}, {Caldwell}, {Deluca}, {Gates}, {Golub}, {Park},
  {Podgorski}, {Bush}, {Scherrer}, {Gummin}, {Smith}, {Auker}, {Jerram},
  {Pool}, {Soufli}, {Windt}, {Beardsley}, {Clapp}, {Lang}, \&
  {Waltham}}]{Lemen12}
{Lemen}, J.~R., {Title}, A.~M., {Akin}, D.~J., {et~al.} 2012, \solphys, 275, 17

\bibitem[{{L{\"o}hner-B{\"o}ttcher} \&
  {Schlichenmaier}(2013)}]{Lohner-Bottcher13}
{L{\"o}hner-B{\"o}ttcher}, J. \& {Schlichenmaier}, R. 2013, \aap, 551, A105

\bibitem[{{Machado} {et~al.}(1980){Machado}, {Avrett}, {Vernazza}, \&
  {Noyes}}]{Machadoetal:1980}
{Machado}, M.~E., {Avrett}, E.~H., {Vernazza}, J.~E., \& {Noyes}, R.~W. 1980,
  \apj, 242, 336

\bibitem[{{Mandrini} {et~al.}(2015){Mandrini}, {Baker}, {D{\'e}moulin},
  {Cristiani}, {van Driel-Gesztelyi}, {Vargas Dom{\'\i}nguez}, {Nuevo},
  {V{\'a}squez}, \& {Pick}}]{Mandrini15}
{Mandrini}, C.~H., {Baker}, D., {D{\'e}moulin}, P., {et~al.} 2015, \apj, 809,
  73

\bibitem[{{Mandrini} {et~al.}(2014){Mandrini}, {Schmieder}, {D{\'e}moulin},
  {Guo}, \& {Cristiani}}]{Mandrini14}
{Mandrini}, C.~H., {Schmieder}, B., {D{\'e}moulin}, P., {Guo}, Y., \&
  {Cristiani}, G.~D. 2014, \solphys, 289, 2041

\bibitem[{{Manini} {et~al.}(2019){Manini}, {Francile}, {L{\'o}pez}, {Hidalgo
  Ramirez}, {Kudaka}, \& {Raulin}}]{Manini19}
{Manini}, F., {Francile}, C., {L{\'o}pez}, F.~M., {et~al.} 2019, Boletin de la
  Asociacion Argentina de Astronomia La Plata Argentina, 61, 38

\bibitem[{{Mauas} {et~al.}(1990){Mauas}, {Machado}, \& {Avrett}}]{Mauas90}
{Mauas}, P. J.~D., {Machado}, M.~E., \& {Avrett}, E.~H. 1990, \apj, 360, 715

\bibitem[{{Meegan} {et~al.}(2009){Meegan}, {Lichti}, {Bhat}, {Bissaldi},
  {Briggs}, {Connaughton}, {Diehl}, {Fishman}, {Greiner}, {Hoover}, {van der
  Horst}, {von Kienlin}, {Kippen}, {Kouveliotou}, {McBreen}, {Paciesas},
  {Preece}, {Steinle}, {Wallace}, {Wilson}, \& {Wilson-Hodge}}]{Meegan09}
{Meegan}, C., {Lichti}, G., {Bhat}, P.~N., {et~al.} 2009, \apj, 702, 791

\bibitem[{{Miteva} {et~al.}(2016){Miteva}, {Kaufmann}, {Cabezas}, {Cassiano},
  {Fernandes}, {Freeland}, {Karlick{\'y}}, {Kerdraon}, {Kudaka}, {Luoni},
  {Marcon}, {Raulin}, {Trottet}, \& {White}}]{Miteva16}
{Miteva}, R., {Kaufmann}, P., {Cabezas}, D.~P., {et~al.} 2016, \aap, 586, A91

\bibitem[{{Nita} {et~al.}(2016){Nita}, {Hickish}, {MacMahon}, \&
  {Gary}}]{Nitaetal:2016}
{Nita}, G.~M., {Hickish}, J., {MacMahon}, D., \& {Gary}, D.~E. 2016, Journal of
  Astronomical Instrumentation, 5, 1641009

\bibitem[{{Ohki} \& {Hudson}(1975)}]{OhkiHudson:1975}
{Ohki}, K. \& {Hudson}, H.~S. 1975, \solphys, 43, 405

\bibitem[{{Penn} {et~al.}(2016){Penn}, {Krucker}, {Hudson}, {Jhabvala},
  {Jennings}, {Lunsford}, \& {Kaufmann}}]{Pennetal:2016}
{Penn}, M., {Krucker}, S., {Hudson}, H., {et~al.} 2016, \apjl, 819, L30

\bibitem[{{Pesnell} {et~al.}(2012){Pesnell}, {Thompson}, \&
  {Chamberlin}}]{Pesnell12}
{Pesnell}, W.~D., {Thompson}, B.~J., \& {Chamberlin}, P.~C. 2012, \solphys,
  275, 3

\bibitem[{{Rast} {et~al.}(2021){Rast}, {Bello Gonz{\'a}lez}, {Bellot Rubio},
  {Cao}, {Cauzzi}, {Deluca}, {de Pontieu}, {Fletcher}, {Gibson}, {Judge},
  {Katsukawa}, {Kazachenko}, {Khomenko}, {Landi}, {Mart{\'\i}nez Pillet},
  {Petrie}, {Qiu}, {Rachmeler}, {Rempel}, {Schmidt}, {Scullion}, {Sun},
  {Welsch}, {Andretta}, {Antolin}, {Ayres}, {Balasubramaniam}, {Ballai},
  {Berger}, {Bradshaw}, {Campbell}, {Carlsson}, {Casini}, {Centeno}, {Cranmer},
  {Criscuoli}, {Deforest}, {Deng}, {Erd{\'e}lyi}, {Fedun}, {Fischer},
  {Gonz{\'a}lez Manrique}, {Hahn}, {Harra}, {Henriques}, {Hurlburt}, {Jaeggli},
  {Jafarzadeh}, {Jain}, {Jefferies}, {Keys}, {Kowalski}, {Kuckein}, {Kuhn},
  {Kuridze}, {Liu}, {Liu}, {Longcope}, {Mathioudakis}, {McAteer}, {McIntosh},
  {McKenzie}, {Miralles}, {Morton}, {Muglach}, {Nelson}, {Panesar}, {Parenti},
  {Parnell}, {Poduval}, {Reardon}, {Reep}, {Schad}, {Schmit}, {Sharma},
  {Socas-Navarro}, {Srivastava}, {Sterling}, {Suematsu}, {Tarr}, {Tiwari},
  {Tritschler}, {Verth}, {Vourlidas}, {Wang}, {Wang}, {NSO and DKIST Project},
  {DKIST Instrument Scientists}, {DKIST Science Working Group}, \& {DKIST
  Critical Science Plan Community}}]{Rast21}
{Rast}, M.~P., {Bello Gonz{\'a}lez}, N., {Bellot Rubio}, L., {et~al.} 2021,
  \solphys, 296, 70

\bibitem[{{Scherrer} {et~al.}(2012){Scherrer}, {Schou}, {Bush}, {Kosovichev},
  {Bogart}, {Hoeksema}, {Liu}, {Duvall}, {Zhao}, {Title}, {Schrijver},
  {Tarbell}, \& {Tomczyk}}]{Scherrer12}
{Scherrer}, P.~H., {Schou}, J., {Bush}, R.~I., {et~al.} 2012, \solphys, 275,
  207

\bibitem[{{Schlichenmaier} {et~al.}(2019){Schlichenmaier}, {Bellot Rubio},
  {Collados}, {Erdelyi}, {Feller}, {Fletcher}, {Jurcak}, {Khomenko},
  {Leenaarts}, {Matthews}, {Belluzzi}, {Carlsson}, {Dalmasse}, {Danilovic},
  {G{\"o}m{\"o}ry}, {Kuckein}, {Manso Sainz}, {Martinez Gonzalez},
  {Mathioudakis}, {Ortiz}, {Riethm{\"u}ller}, {Rouppe van der Voort}, {Simoes},
  {Trujillo Bueno}, {Utz}, \& {Zuccarello}}]{EST19}
{Schlichenmaier}, R., {Bellot Rubio}, L.~R., {Collados}, M., {et~al.} 2019,
  arXiv e-prints, arXiv:1912.08650

\bibitem[{{Sim{\~o}es} {et~al.}(2017){Sim{\~o}es}, {Kerr}, {Fletcher},
  {Hudson}, {Gim{\'e}nez de Castro}, \& {Penn}}]{Simoesetal:2017}
{Sim{\~o}es}, P. J.~A., {Kerr}, G.~S., {Fletcher}, L., {et~al.} 2017, \aap,
  605, A125

\bibitem[{{Sim{\~o}es} {et~al.}(2019){Sim{\~o}es}, {Reid}, {Milligan}, \&
  {Fletcher}}]{Simoesetal:2019A}
{Sim{\~o}es}, P. J.~A., {Reid}, H. A.~S., {Milligan}, R.~O., \& {Fletcher}, L.
  2019, \apj, 870, 114

\bibitem[{{Trottet} {et~al.}(2015){Trottet}, {Raulin}, {Mackinnon},
  {Gim{\'e}nez de Castro}, {Sim{\~o}es}, {Cabezas}, {de La Luz}, {Luoni}, \&
  {Kaufmann}}]{Trottet15}
{Trottet}, G., {Raulin}, J.~P., {Mackinnon}, A., {et~al.} 2015, \solphys, 290,
  2809

\bibitem[{{Turon} \& {L{\'e}na}(1970)}]{Turon70}
{Turon}, P.~J. \& {L{\'e}na}, P.~J. 1970, \solphys, 14, 112

\bibitem[{{van Driel-Gesztelyi} {et~al.}(2012){van Driel-Gesztelyi}, {Culhane},
  {Baker}, {D{\'e}moulin}, {Mandrini}, {DeRosa}, {Rouillard}, {Opitz},
  {Stenborg}, {Vourlidas}, \& {Brooks}}]{vanDriel12}
{van Driel-Gesztelyi}, L., {Culhane}, J.~L., {Baker}, D., {et~al.} 2012,
  \solphys, 281, 237

\bibitem[{{van Driel-Gesztelyi} \& {Green}(2015)}]{vanDriel15}
{van Driel-Gesztelyi}, L. \& {Green}, L.~M. 2015, Living Reviews in Solar
  Physics, 12, 1

\bibitem[{{Vernazza} {et~al.}(1981){Vernazza}, {Avrett}, \& {Loeser}}]{VAL81}
{Vernazza}, J.~E., {Avrett}, E.~H., \& {Loeser}, R. 1981, \apjs, 45, 635

\bibitem[{{White} {et~al.}(2011){White}, {Benz}, {Christe}, {F{\'a}rn{\'\i}k},
  {Kundu}, {Mann}, {Ning}, {Raulin}, {Silva-V{\'a}lio}, {Saint-Hilaire},
  {Vilmer}, \& {Warmuth}}]{White11}
{White}, S.~M., {Benz}, A.~O., {Christe}, S., {et~al.} 2011, \ssr, 159, 225

\bibitem[{{Woods} {et~al.}(2012){Woods}, {Eparvier}, {Hock}, {Jones},
  {Woodraska}, {Judge}, {Didkovsky}, {Lean}, {Mariska}, {Warren}, {McMullin},
  {Chamberlin}, {Berthiaume}, {Bailey}, {Fuller-Rowell}, {Sojka}, {Tobiska}, \&
  {Viereck}}]{Woods12}
{Woods}, T.~N., {Eparvier}, F.~G., {Hock}, R., {et~al.} 2012, \solphys, 275,
  115

\end{thebibliography}


\end{document}